\documentclass[pre,amsmath,twocolumn]{revtex4-1}
\usepackage[normalem]{ulem}
\usepackage{epsfig}
\usepackage{amsfonts} 
\usepackage{amsmath} 
\usepackage{nicefrac}
\usepackage{color} 
\usepackage{hyperref}
\setcitestyle{square}
\usepackage{subfigure} 
\usepackage{natbib} 
\usepackage{graphicx,amssymb}

\newcommand{\red}{\textcolor{red}}

\newcommand{\ignore}[1]{} 

\def\eq#1{${#1}$} 
\def\EQ#1#2{\begin{equation}{#1}\label{#2}\end{equation}} 
 
\newcommand{\Tau}{\tau^*}
\newcommand{\Intd}{{\rm{d}}} 
 
\newcommand{\Hquad}{\hspace{0.5em}} 
\newcommand{\HHquad}{\hspace{0.25em}} 
\usepackage{mathrsfs}
\usepackage{commath} 
\newcommand{\OO}{{\mathcal{O}}}

\newcommand{\MeanV}{\mathbf{\mathsf{v}}} 
\usepackage{gensymb} 
\usepackage{mathtools} 
\usepackage[makeroom]{cancel}

\setcitestyle{square}
\usepackage{subfigure}

\usepackage{enumitem}
\usepackage{lipsum}

\begin{document} 

\title{Moses, Noah and Joseph Effects in  L\'evy Walks}
\author{Erez Aghion\textsuperscript{a},  Philipp G. Meyer\textsuperscript{a}, Vidushi Adlakha\textsuperscript{b,c}, Holger Kantz\textsuperscript{a}, Kevin E.  Bassler\textsuperscript{b,c,d}}
 \affiliation{\makebox[\textwidth][c]{a) Max-Planck Institute for the Physics of Complex Systems, Dresden D-01187, Germany} \\{ b) Department of Physics, University of Houston, Houston, TX 77204, USA } \\ {c) Texas Center for Superconductivity, University of Houston, Houston, TX 77204, USA } \\ {d)Department of Mathematics, University of Houston, Houston, TX 77204, USA}}
%
%
\begin{abstract}
We study a method for detecting the origins of anomalous diffusion, when it is observed in an ensemble of times-series, generated  experimentally or numerically, without having knowledge about the exact underlying dynamics. The reasons for anomalous diffusive scaling of the mean-squared displacement are decomposed into three root causes: increment correlations are expressed by the ``Joseph effect" \cite{mandelbrot1968noah}, fat-tails of the increment probability density lead to a ``Noah effect" \cite{mandelbrot1968noah}, and non-stationarity, to the ``Moses effect" \cite{chen2017anomalous}. 
After appropriate rescaling, based on the quantification of these effects, the increment distribution converges at increasing times to a time-invariant asymptotic shape. For different processes, this asymptotic limit can be an equilibrium state, an infinite-invariant, or an infinite-covariant density. We use numerical methods of time-series analysis to quantify the three effects in a model of a non-linearly coupled L\'evy walk,  compare our results to theoretical predictions, and discuss the generality of the method. 
 
\end{abstract}

\maketitle{}



\section{Introduction} 
\label{Introduction}

Diffusive processes that scale anomalously with time, such that the Mean-Squared Displacement (MSD) of the expanding particle packet is
\begin{equation}
    \langle x^2(t)\rangle \sim t^{2H},
\label{HurstDefinition}
\end{equation} 
and the Hurst exponent $H\neq1/2$, are widely observed.
This behavior is found both in theoretical models as well as in many experiments, see e.g.  \cite{hofling2013anomalous,metzler2014anomalous,metzler2019brownian,oliveira2019anomalous,sabri2020elucidating}. Of-course, if we know the exact  underlying process responsible for the dynamics, \eq{H} can be determined exactly and the various features of the system that lead to the deviation from the standard linear scaling of the MSD, expected by the Gaussian Central Limit Theorem (CLT), can be understood. However, when anomalous diffusive scaling is detected in measurements it is not always clear what is responsible for the observed behavior of the system. Imagine, for example, that we obtain an ensemble of data-series describing intra-day trades in financial markets~\cite{bassler2007nonstationary,seemann2012ensemble,chen2017anomalous}, or experimental data obtained from observation of molecules diffusing inside cells, e.g.,  \cite{tolic2004anomalous,brauchle2010single,xie2008single,weigel2011ergodic,krapf2019spectral,sabri2020elucidating}. Here, the proper characterization of the exact root causes of this phenomenon is very important, since it can have implications on how we understand the underlying functioning of the system. 
If, for example, we observe that the MSD grows faster than linearly with time, is this due to temporal correlations in the data that cause random large fluctuations to be followed by similar or even greater ones?  Is it the result of a fat-tailed increment distribution, or is it because there is an actual trend of inflation in the system? Our analysis below allows us to give answer to these questions, despite the fact that we cannot completely restore the underlying process just from the data. 

To make this more precise: Consider a continuous-time stochastic process \eq{x(t')} defined in the time interval \eq{t'\in[0,t]}. We can choose a number \eq{Q} of observation windows of duration \eq{\Delta=t/Q}, and then, represent this process by a discrete time-series composed of consecutive \emph{increments}, starting at times \eq{\left\{0,\Delta,2\Delta,\ldots,(Q-1)\Delta\right\}}. The increments are $\left\{\delta x_{1},\delta x_{2}, \ldots,\delta_{Q}\right\}=\left\{x(\Delta)-x(0),x(2\Delta)-x(\Delta),\ldots,x(t)-x(t-\Delta)\right\}$. According to the Gaussian CLT, in the limit of large \eq{Q}, if the increments are independent, identically distributed (IID) random variables chosen from a distribution with finite variance, then the MSD will grow linearly with \eq{Q} and thus with time.
Each of the three ways that the CLT can be violated corresponds to a \emph{constitutive effect} that can produce anomalous scaling~\cite{chen2017anomalous}. 
For processes with stationary increments, where the probability distribution of \eq{\delta x_j} is independent of time,  anomalous diffusive scaling can occur because of long-time increment correlations. This is called the \emph{Joseph effect}~\cite{mandelbrot1968noah,chen2017anomalous,meyer2018anomalous}. A paradigmatic process that exhibits this effect is fractional Brownian motion \cite{lim2002self,chen2017anomalous}. Another cause of anomalous scaling may be that the increment distribution is fat-tailed, in the sense that its second moment is divergent. This is the \emph{Noah effect} \cite{mandelbrot1968noah,chen2017anomalous,meyer2018anomalous}. A L\'evy flight process where the increments are power-law distributed, independent random variables~\cite{shlesinger1986levy,metzler2014anomalous}, but with infinite variance, is one example of a model with this effect. 
 When the increment distribution is non-stationary, anomalous diffusive scaling can also arise due to the \emph{Moses effect} \cite{chen2017anomalous,meyer2018anomalous}. A paradigmatic model in this case is scaled Brownian motion \cite{jeon2014scaled,thiel2014scaled,safdari2015aging}.  
 Each of the three effects can appear individually in a system, or in various combinations. Importantly, the three effects can be interconnected with each-other. For example, in \cite{meyer2018anomalous}, it was shown that statistical aging in the process can be associated not only with a Moses, but Noah effect. Among other things, this manuscript will extend our understanding of the coupling of the Moses and Noah effects. 
 The quantification of the three constitutive effects and the relation between them is given in Sec. \ref{SecThreeExps}.  

In this manuscript, we investigate these three constitutive effects in a well studied stochastic process called coupled L\'evy walk~\cite{LevyWalks}. This model is known to have a rich spectrum of statistical behaviors, found by the tuning of a few well defined handles. We explore the emergence of the three effects in different parameter regimes of the model using simulations and methods of time-series analysis of single L\'evy walk trajectories, and compare our findings with analytical results based on the well developed theory for this process. This example shows that the analysis based on the three constitutive effects is a useful tool that can be applied to study other systems as well (see discussion).    

In a two-state L\'evy walk  \cite{shlesinger1986levy,LevyWalks, froemberg2015asymptotic}, a particle starts at \eq{x=0} at time \eq{t'=0} and then moves in independent steps. Each step has a random duration $\tau$, chosen from a Probability Density Function (PDF) of the form
\EQ{g(\tau)\sim \frac{c}{|\Gamma(-\gamma)|}\tau^{-1-\gamma}}{gTauDefinition} 
at long \eq{\tau}, where \eq{c,\gamma>0} are constants. 
During each step, the particle travels at a constant velocity \eq{V}, whose magnitude \eq{|V|} can be either $\pm1$ (sometimes referred to as ``genuine L\'evy walk" \cite{LevyWalks}), or a deterministic function of \eq{\tau}, but whose direction is chosen randomly to be either toward the right, along the positive \eq{\hat{x}-}axis (\eq{+}), or left (\eq{-}) along the negative axis. The latter, generalized model, is the case studied in detail in this manuscript (see also e.g., \cite{shlesinger1987levy,akimoto2013distributional,akimoto2014phase,albers2014weak,aghion2018asymptotic}), and the results include also the constant-velocity case.  The probability of the direction being to the right or to the left is equal, so the motion is unbiased and the velocity has a symmetric PDF \eq{\phi(V)}.
At time \eq{t'=t}, the process stops. Up to this point, the particle has made \eq{N-1} complete steps, and one, final ``partial" step of duration  \EQ{\Tau=t-\sum_{i=1}^{N-1}\tau_i.}{TotalTime} 
The properties of the final step have been shown to have a dramatic affect on the overall behavior of the system \cite{LevyWalks}, as the velocity \eq{V_N} during this step  does not necessarily have to be distributed like all its predecessors, see e.g., \cite{froemberg2015asymptotic}.  For more on this point, see Sec. \ref{SecModel}.  
The number of steps in the process \eq{N \in [1,\infty)}, in the time interval $[0,t]$ is random, and the particle's position at time \eq{t} is given by the sum \eq{x(t)=\sum_{i=0}^{N-1}\chi_i+\chi^*}, where  \eq{\chi_i=V_i\tau_i}, and \eq{\chi^*=V_N\tau^*}. 

Table \ref{TableNotations} summarizes the main notations we use throughout the paper, by order of their appearance in the main text. 
\begin{table}[t] 
\footnotesize
\centering 
\begin{tabular}{c c } 
\hline\hline 
Notation& Definition\\
[0.5ex] %
\hline 
$V,\tau,\chi$&L\'evy walk: step- velocity, duration, displacement\\
$M,L,J,H$&Exponents: Moses, Noah, Joseph, Hurst\\
$\Delta,\delta x$&Time series: increment- duration, size \\
$\MeanV$&Time series: Mean velocity during an increment\\
$\alpha,\beta$&Exponents describing the  shape of the distribution of $\MeanV$\\
 $z_\beta$& $\MeanV/t^\beta$\\
$v(t)$&Instantaneous velocity of the L\'evy walker at time \eq{t}\\
$\tilde{v}$& $v/t^{\nu-1}$\\ 
[0.5ex] %
\hline\hline 
\end{tabular} 
\caption{\footnotesize{The main notations used in this manuscript, by order of their appearance in the main text. Note that the instantaneous velocity $v(t)$ is not always defined, for example in the case of Brownian motion. This does not matter for the general analysis of the three effects, which are defined via  $\MeanV$, see Sec. \ref{SecThreeExps}. In the example that we use to demonstrate our analysis, namely L\'evy walk, $v(t)$ exists, and  $\lim_{\Delta\rightarrow0}\MeanV\rightarrow v(t)$, see also Sec. \ref{SecMAndL}.}}
\label{TableNotations} 
\end{table} 
The structure of the manuscript is as follows: In Sec. \ref{SecThreeExps}, we define the three exponents that quantify the Moses, Noah and Joseph effects. We discuss the relation between them, and their role in determining the scaling shape of the increment PDF. In Sec. \ref{SecModel}, we extend the details on the  L\'evy walk model. In Sec. \ref{SecMainResults}, we provide a summary of our main results, obtained from  time-series analysis of numerical simulations, and a brief comparison of these results with the theoretical predictions.  In Sec. \ref{SecMAndL} we obtain analytic results for the Moses and Noah effects, and in Sec. \ref{JosephExponent} for the Joseph. 
We generalize the model in Sec. \ref{Section5}, and the discussion is provided in Sec. \ref{Discussion}.

 \section{Story of three exponents:~\eq{M, L} and~\eq{J}} 
 \label{SecThreeExps}

The complete decomposition of the origin of anomalous diffusion presented in the introduction, was originally derived for discrete-time processes \cite{chen2017anomalous}. In this case, the process starts at \eq{\xi_0=0}, at \eq{n=0}, and evolves in  discrete jumps \eq{n=1...N} with  duration \eq{\Delta}, until time \eq{t=N\Delta}. The particle's position after \eq{n} steps is denoted \eq{\xi_{n\Delta}}. The Moses effect is quantified by the exponent \eq{M}, given by the median of the sum, of the absolute value of the time-series increments~\cite{chen2017anomalous} $m\left[\sum_{n=1}^{t/\Delta} |\delta \xi_n|\right]\equiv m\left[\sum_{n=1}^{t/\Delta}|\xi_{n\Delta}-\xi_{(n-1)\Delta}|\right]\propto    t^{M+1/2}$. Here, \eq{M=1/2} yields a linear relation which is similar to normal diffusion. The Noah effect is defined by the scaling of the median of the sum of square-increments, and quantified by the Latent exponent $L$: $ m\left[\sum_{n=1}^{t/\Delta} (\delta \xi_n)^2\right]\propto    t^{2L+2M-1}$. Here again, normal diffusion leads to linear scaling, where $M=L=1/2$. If there is no Moses effect, namely \eq{M=1/2}, the deviation from this scaling is quantified only by the exponent \eq{L}, and it  arises if the increment PDF is fat-tailed.
Finally, the Joseph exponent can be defined via the sum over the auto-correlation function \cite{meyer2018anomalous}   $\sum_{\Delta'=0}^{\tilde{\Delta}}\langle \delta \xi_n\delta \xi_{n+{\Delta'}}\rangle/\langle (\delta \xi_n)^2\rangle\propto \tilde{\Delta}^{2J-1}$, where \eq{0\leq J\leq 1}. {Here, starting from an arbitrary time point \eq{n}, we sum over a discrete lag time \eq{\Delta'}, up to e.g., \eq{\tilde{\Delta}\sim \OO(t/10)} (\eq{\tilde{\Delta}} is not related to \eq{\Delta}, defined above), and the scaling shape is valid when  \eq{\tilde{\Delta},t\gg1}. When \eq{J>1/2}, the correlations decay very slowly with \eq{\tilde{\Delta}}, which leads to a divergent sum when \eq{\tilde{\Delta}\rightarrow\infty},  and superdiffusion (see discussion on ``long-ranged correlations" e.g., in \cite{Beran}). When \eq{J\leq1/2}, the correlation function decays at least as fast as \eq{1/\tilde{\Delta}}, which may lead either to normal diffusion, or in some particular cases to sub-diffusion, see Appen. \ref{AppenHVsMLJ}}. 

\begin{figure}[t]
\centering
\includegraphics[width=0.5\textwidth]{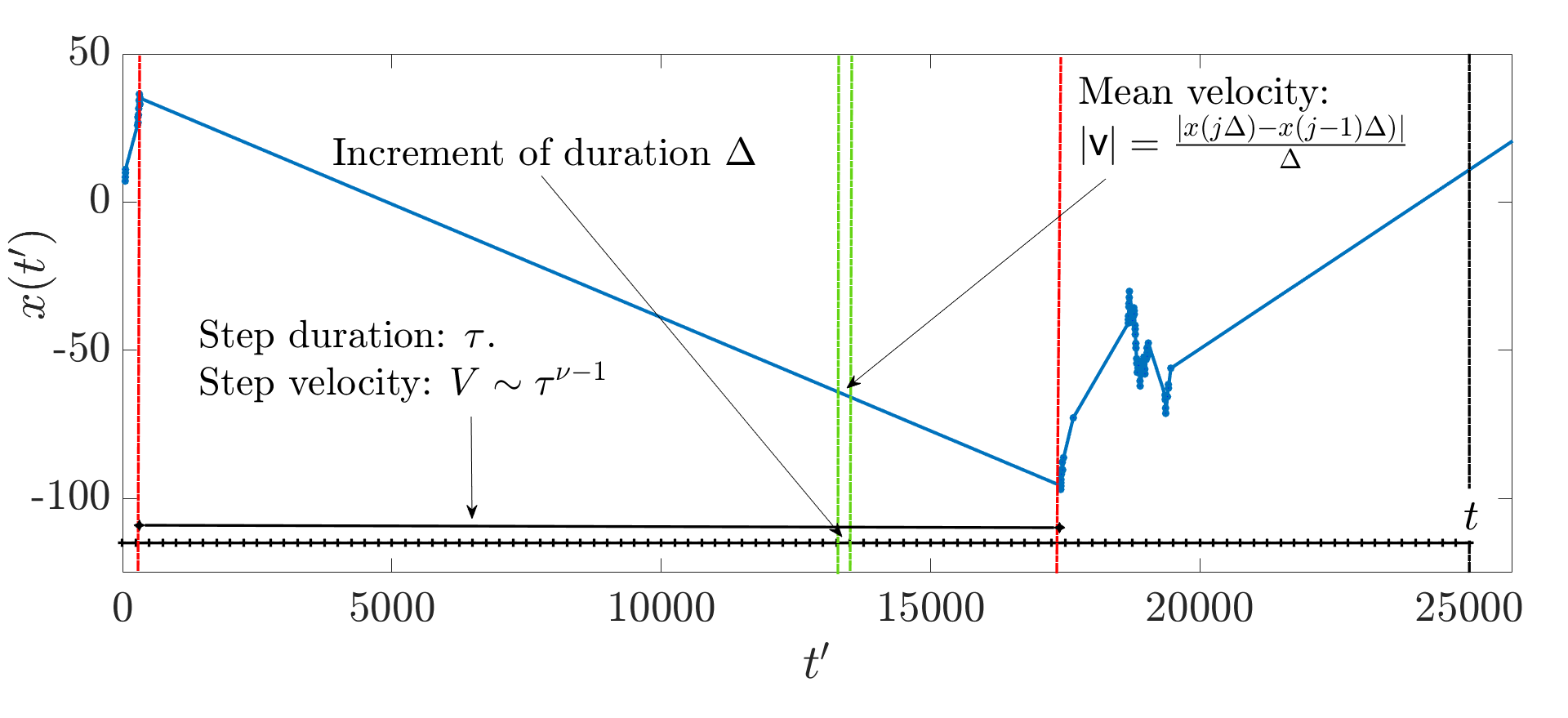}
 \caption {\footnotesize{An example of a L\'evy walk path $x(t')$ (blue) versus time, generated by the model in Sec. \ref{SecModel}. At the total measurement time $t$, the last step is incomplete. Two red dash-dot lines mark the start and end points of one completed L\'evy walk step, whose duration $\tau$ was selected from the PDF Eq. \eqref{MLDistributionOfTaus}, and the step-velocity is $V\sim\tau^{\nu-1}$. Here $\gamma=0.52, \nu=0.5$. 
 As explained in Sec. \ref{SecThreeExps}, the trajectory is decomposed into a series of consecutive increments $n=1,2,...$, of equal duration $\Delta$, the start and end points of one such increment are marked e.g. by two green dash-dot lines. The size of the average velocity  $|\MeanV|$ in that increment is also presented.}}
 \label{FigCTRWToTSA}
\end{figure}

For a process \eq{x(t)} in continuous time, we divide the time series into \eq{Q} non-overlapping observation windows of duration \eq{\Delta=t/Q} as mentioned in the introduction, and define the average velocity in each time interval 
 $\MeanV(t')\equiv| \delta x_j|/\Delta$, where $\delta x_j=x(j\Delta)-x\left[(j-1)\Delta\right]$, and  \eq{(j-1)\Delta<t'<j\Delta}.  Fig. \ref{FigCTRWToTSA} illustrates the decomposition of a continuous-time random trajectory, into a time-series of $N$  increments of equal duration $\Delta\ll t$. Now, we can re-write the definition of the Moses effect in terms of the ensemble-time averaged absolute-velocity 
 (when \eq{\Delta\ll t}) 
  \begin{equation}
     \left\langle\overline{|\MeanV|}\right\rangle\equiv\left\langle\frac{1}{t-\Delta}\sum_{j=1}^{t/\Delta}\frac{|\delta x_j|}{\Delta}\right\rangle\propto t^{M-1/2}.
 \label{MosesDefinitionContinousTime}  \end{equation} 
 We use here the ensemble mean, instead of the median, since it is a more convenient property to study analytically and numerically, hence we assume by this definition that this mean does not diverge. 
In the same spirit, the Noah effect is defined via the ensemble-time average of the squared velocity, when \eq{\Delta\ll t} 
\begin{align}
     &\left\langle\overline{\MeanV^2}\right\rangle\equiv\left\langle\frac{1}{t-\Delta}\sum_{j=1}^{t/\Delta} \frac{\left(\delta x_j\right)^2}{\Delta^2}\right\rangle\propto  t^{2L+2M-2}, 
      \label{NoahDefinitionContinousTime}
  \end{align} 
where  
\begin{equation} 
1/2\leq L\leq 1. 
\label{LatentLimits}
\end{equation} 
In this definition, one can notice that manifestation of the Noah effect is somewhat different from the case of e.g., a L\'evy flight, since the mean of the squared increments is not divergent. In fact, as we explain in detail below, what leads to $L\neq1/2$ in this case, is that the increment PDF has a regime where its shape is fat-tailed, but this regime has a time-dependent cutoff which is pushed towards $\pm\infty$ as time increases. The resemblance between this observation, and the source of  the Noah effect in its original definition on P. $1$, is the reason that we can make the association between the two cases and refer to $L$ in Eq. (\ref{NoahDefinitionContinousTime},\ref{LatentLimits}) throughout this manuscript as the Latent exponent.
 The upper bound on $L$, in Eq. \eqref{LatentLimits}, is true because $\langle\overline{\MeanV^2}\rangle\leq \langle\overline{|\MeanV|}\rangle^2$. Intuitively it means that a tuning of the parameter that leads to a Noah effect beyond $L=1$, would automatically increase the scaling exponent of the first moment and therefore lead to aging and a Moses effect, instead of Noah. The lower bound exists  because fat tails of the increment distribution, which are described by a Noah effect, can never lead to a slowing down of the process.

In this work, we will assume that also $\left\langle{|\MeanV|}\right\rangle\propto  t^{M-1/2}$  and $\left\langle{\MeanV^2}\right\rangle\propto  t^{2L+2M-2}$. We address the relation between our definitions and the original time-averaged definitions of these effects, which were derived when the ensemble means could be divergent, below (Sec. \ref{Scaling_shapes}). 
Since the ensemble and time averaging procedures are commutative, if we know the first we can immediately obtain the latter via $\left\langle\overline{|\MeanV|}\right\rangle\rightarrow\overline{\langle|\MeanV|\rangle}=(1/t)\int_0^t\langle|\MeanV|(t')\rangle\Intd t'$ which yields \eq{=(Const./t)\int_0^t {t'}^{M-1/2}\Intd t'=[1/(M+1/2)]\langle|\MeanV|\rangle}. Since we can find \eq{\left\langle\overline{\MeanV^2}\right\rangle} in a similar way from its ensemble mean, this yields  
\begin{align} 
\left\langle\overline{|\MeanV|}\right\rangle=\frac{\langle |\MeanV|\rangle}{M+1/2},\Hquad\mbox{and}\Hquad
\left\langle\overline{\MeanV^2}\right\rangle=\frac{\langle \MeanV^2\rangle}{2L+2M-1}.
\label{EATAVsEA} 
\end{align} 
Note that Eq. \eqref{EATAVsEA} introduces additional limits on the possible values of $M$ and $L$, for processes with finite $\langle|\MeanV|\rangle$ and $\langle\MeanV^2\rangle$, since the ratio between the time and ensemble averages here has to be positive. These limits are consistent with our results for the L\'evy walk model, in Sec.~\ref{SecMainResults}. 
  
  We define the Joseph exponent also in the spirit of the discrete case, via the scaling of the integral 
 \eq{\int_0^{\tilde{\Delta}} d\Delta^\prime \langle \MeanV(t)\MeanV(t+\Delta')\rangle/\langle \MeanV^2\rangle\propto \tilde{\Delta}^{2J-1},} 
 for large $\tilde{\Delta}$. Here again,  \eq{\tilde{\Delta}} should not be confused with \eq{\Delta}, which is the time duration from which we defined \eq{\MeanV}. 
 In this manuscript we will only focus on the case where \EQ{1/2\leq J\leq 1,}{JRange} see Appen. \ref{AppenHVsMLJ} for more explanation. Taking the derivative of the integral with respect to \eq{{\tilde{\Delta}}}, the autocorrelation function is 
 \begin{equation}
     f({\tilde{\Delta}})\equiv\frac{\left\langle \MeanV(t)\MeanV(t+{\tilde{\Delta}})\right\rangle}{\langle \MeanV^2(t)\rangle}\propto {\tilde{\Delta}}^{2J-2}, 
 \label{JosephDefinitionContinousTime}
 \end{equation} 
 at \eq{{\tilde{\Delta}}\gg 1}. For small $\tilde{\Delta}$, we define  \eq{f({\tilde{\Delta}})\equiv f_<({\tilde{\Delta}})}, where \eq{f_<} insures that the autocorreletation function is regularized at \eq{{\tilde{\Delta}}\rightarrow0}. 
Note that in data analysis there are several known methods to obtain the Joseph exponent   without directly calculating the autocorrelation function. These methods have various advantages and disadvantages in practice, see Sec. \ref{JosephExponent} and Appen. \ref{Appenctamsd} and \ref{DFAAndEATMSD}.  

We note that by dividing $\delta x_j$ by \eq{\Delta}, and defining the three effects via the mean increment velocity $\MeanV$, we did not limit the generality of the definitions at all. The reason is that we did not at this point take the limit \eq{\Delta\rightarrow0}, hence we do not require the instantaneous velocity to be defined. In any process, one can discuss average velocities and increments of a finite-time duration interchangeably. 
 
\subsection{ Relation between $M,L,J$ and $H$}  
\label{SecRelationMLJH} 
Let \eq{\MeanV(0)\equiv0}, using Eq. \eqref{JosephDefinitionContinousTime} and the Green-Kubo relation \cite{meyer2017greenkubo}, the MSD of the process can be written as   
  \small 
  \begin{align}
      &\langle x^2\rangle=2\int_0^t\Intd {\tilde{\Delta}}\int_0^{t-{\tilde{\Delta}}}\Intd t' \langle \MeanV(t')\MeanV(t'+{\tilde{\Delta}})\rangle\nonumber\\ 
      &\propto 2\int_1^t\Intd {\tilde{\Delta}}\int_0^{t-{\tilde{\Delta}}}\Intd t' \langle  \MeanV^2(t')\rangle{\tilde{\Delta}}^{2J-2}\nonumber\\ 
      &+2\int_0^1 \Intd {\tilde{\Delta}} f_<({\tilde{\Delta}})\int_0^t\langle v^2(t')\rangle\Intd t'\nonumber\\ 
      &\propto 2\int_1^t\Intd {\tilde{\Delta}}{\tilde{\Delta}}^{2J-2}\int_0^{t-{\tilde{\Delta}}}\Intd t' {t'}^{2L+2M-2}+c_<t^{2L+2M-1}\nonumber\\ 
      &\propto\frac{2}{2L+2M-1}\int_1^t\Intd {\tilde{\Delta}} {\tilde{\Delta}}^{2J-2}(t-{\tilde{\Delta}})^{2L+2M-1}+c_<t^{2L+2M-1}\nonumber\\
      &\underbrace{\sim}_{\footnotesize{\begin{aligned}u&\leftrightarrow{\tilde{\Delta}}/t \\ &t\rightarrow\infty\end{aligned}}}\frac{2t^{2L+2M+2J-2}}{2L+2M-1}\int_0^1\Intd u u^{2J-2}(1-u)^{2L+2M-1}\nonumber\\
      & \qquad\qquad \propto t^{2L+2M+2J-2},\qquad\mbox{when}\qquad J>1/2. 
      \label{ConnectionBetweenExponents1}
  \end{align}
\normalsize 
In Eq. \eqref{ConnectionBetweenExponents1}, \eq{c_<} is a constant, and  in the last step note that since the term \eq{\propto t^{2M+2L-1}} is subdominant with respect to the other when \eq{J>1/2}, we neglected it in the long-time limit. Using Eq. \eqref{HurstDefinition}, this yields 
\begin{equation}
  H=J+L+M-1.  
  \label{hjlm}
\end{equation}
The relation in Eq. \eqref{hjlm} was previously shown to hold empirically in a number of models in \cite{chen2017anomalous,meyer2018anomalous}. It was conjectured to be broadly valid, even for systems beyond the case we study here, in particular also when ensemble averages diverge and the Moses and Noah effects are only quantified via their original time-averaged definitions. However, a rigorous derivation in other cases is still needed. For more details see Appen.~\ref{AppenHVsMLJ}.

\subsection{Scaling shapes of the increment distribution}
\label{Scaling_shapes}

Considering ensemble averages allows us to obtain additional insight about the meaning of the Moses and Noah effects. Assume that \eq{\langle|\MeanV|\rangle} and \eq{\langle \MeanV^2\rangle} are not divergent. Let $P_t(\MeanV)$ be the  PDF of finding an increment velocity  $\MeanV$ at time $t$, given that the process started at at rest at $t = 0$. This increment PDF is said to have a single scaling shape, if for any $x$ and $t$ it can be described by a time-independent function $W(z_\beta)$, such $W(z_{\beta})=t^{\beta}P(\MeanV/t^{\beta})$ and $z_{\beta}=\MeanV/t^{\beta}$. In our case, we do not restrict $P_t(\MeanV)$ to only one such scaling regime, and it can have two different scaling shapes in it bulk and the tails, a situation not uncommon in anomalous diffusion which is associated with multifractality, see e.g., \cite{castiglione1999strong,seuront2014anomalous,rebenshtok2014non,kessler2010infinite,aghion2018asymptotic,grahovac2015asymptotic}. If both the mean of \eq{|\MeanV|} and \eq{\MeanV^2} are taken from the same scaling regime of  \eq{P_t(\MeanV)}, then in this regime   
 \EQ{\lim_{t\rightarrow\infty}t^{\alpha+\beta}P_t(\MeanV/t^\beta)\rightarrow W(z_\beta),\quad\mbox{ where}\quad z_\beta=\MeanV/t^\beta,}{ICD}
 and 
 \begin{align}
 &\qquad\quad L=\alpha/2+1/2,\quad M=\beta-\alpha+1/2\nonumber\\ 
 &(\mbox{equivalently:}\quad \alpha=2L-1,\quad \beta=M+2L-3/2).
 \label{ALphaBetaDefinitions} 
 \end{align} 
 Notice that since \eq{1/2\leq L\leq1}, Eq. \eqref{LatentLimits}, then \eq{0\leq\alpha\leq1}. The limit function \eq{W(z_\beta)} is responsible for the mean of \eq{|\MeanV|} and \eq{\langle \MeanV^2\rangle} via \EQ{\langle|\MeanV|^q\rangle=2\int_{0}^\infty\Intd \MeanV|\MeanV|^q P_t(\MeanV)\underbrace{\approx}_{t\gg1}2t^{q\beta-\alpha}\int_0^\infty \Intd z_\beta|z_{\beta}|^qW(z_\beta),}{ICD1} 
 for \eq{q=1,2}. 
 
 When \eq{M,L} are such that both \eq{\alpha} and \eq{\beta} are zero, the increment PDF has a stationary asymptotic (equilibrium) state. Coincidentally this occurs only when \eq{M=L=1/2}, which as mentioned means that the time-series satisfies at least two of the conditions of the Gaussian CLT. Curiously, \eq{M} can also be half if \eq{\alpha=\beta\neq0}. When \eq{P_t(\MeanV)} is non-stationary, we always have a Moses effect. The PDF has a normalized scaling shape, when  \eq{\alpha=0} but \eq{\beta\neq0}, namely \eq{L=1/2,M\neq1/2}. This is the onset of a ``pure" Moses effect. Now, the exponent \eq{M} tells us how to re-scale the PDF in-order to find the invariant limit, since \eq{P_t(\MeanV)\sim t^{1/2-M}W(\MeanV/t^{-1/2+M})}. According to Eqs. (\ref{ICD},\ref{ICD1}), if we define \eq{\langle |z_\beta|^q\rangle_\mathbb{W}\equiv\int_0^\infty |z_\beta|^qW(z_\beta)\Intd z_\beta} for $q=1,2$, then $\langle |v|^q\rangle=2t^{q(1/2-M)}\langle|z_\beta|^q\rangle_\mathbb{W}$. Note that usually, based on intuition taken from Gaussian processes, there is a tendency to vaguely associate the Hurst exponent \eq{H}, with the "self-similarity" property of the process. However in anomalous diffusion that is not necessarily the case; one example is when the MSD is diverging, e.g., in L\'evy flight, another example is the case of multifractality \cite{castiglione1999strong}. In our case, it is \eq{\beta}, not \eq{H}, that may describe this property, from the point of view of the increment PDF.

 The onset of a Noah effect means that \eq{\MeanV^2} becomes non-integrable with respect to the scaling function which gives the shape of the $P_t(\MeanV)$ in the bulk. 
In the paradigmatic example for this effect, L\'evy flight \cite{mandelbrot1968noah}, the PDF \eq{P_t(\MeanV)} can be e.g., a stationary symmetric L\'evy distribution \eq{l_{\xi,1,0}(\MeanV)}, with $0<\xi<2$, defined as the inverse-Laplace transform of \eq{\exp(-|u|^\xi)}, from \eq{u \rightarrow \MeanV} \cite{klafter2011first}.  In this case, by definition, there is no Moses effect, and the Noah effect rises since $\int_{-\infty}^\infty \MeanV^2 l_{\xi,1,0}(\MeanV)\Intd\MeanV\rightarrow\infty$, though of-course, here it can only be quantified by the original definition of \eq{L}, namely via the time-average of the squared increments of single time-series \cite{mandelbrot1968noah}. If the increment PDF would have e.g., the scaling shape 
$P_t(\MeanV)\sim t^{-1/\xi}l_{\xi,1,0}(\MeanV/t^{1/\xi})$, we would find both a Moses effect, and a Noah effect which is still characterized via the time average.

 A more involved scenario that can occur, is when the large fluctuations of the system are reduced such that \eq{\langle \MeanV^2\rangle} is not strictly infinity, but is increasing with time as in Eq. \eqref{NoahDefinitionContinousTime}, because at its tails the PDF $P_t(\MeanV)$ is scaled differently in time with respect to the bulk. Now, the definitions in Eqs. (\ref{MosesDefinitionContinousTime},\ref{NoahDefinitionContinousTime}) are valid. The Noah effect will now appear if the function which describes the asymptotic shape of $P_t(\MeanV)$ at the bulk is fat-tailed (in the sense that its variance is infinite), but the mean \eq{\langle \MeanV^2\rangle} will be given by a second scaling function to which $P_t(v)$ convergence at the tails. If it happens that the mean of \eq{|v|} and \eq{v^2} are obtained from different scaling regimes, then again Eq. \eqref{ICD} and Eq. \eqref{ALphaBetaDefinitions} are not valid, but one can use methods such as estimating fractional moments \cite{rebenshtok2014non,aghion2018asymptotic,grahovac2015asymptotic} to find the various scaling shapes of \eq{P_t(\MeanV)}. If both $\langle |\MeanV|\rangle$ and $\langle \MeanV^2\rangle$ correspond to the second  scaling function (that describes the large fluctuations), and are proportional to $t^{M-1/2}$ and $t^{2M+2L-2}$ respectively, then Eq. \eqref{ICD} is valid. But in this case, \eq{W(z_\beta)} which denotes these moments might not be normalizable, namely \eq{\int_0^\infty W(z_\beta)\Intd z_\beta\rightarrow\infty}. Here, \eq{\alpha} and the Latent exponent \eq{L} serves as a measure of \textit{how far} the increment PDF is from having a normalized limit shape. When \eq{\alpha>0} and \eq{\beta=0}, equivalently \eq{L>1/2} and \EQ{M=\frac{3}{2}-2L,}{LAndMID} \eq{W(z_\beta)} is an infinite-invariant density, a type of quasi-equilibrium state, see e.g.,  \cite{aaronson1997introduction,korabel2009pesin,leibovich2019infinite,akimoto2019infinite,aghion2019infinite,sato2019anomalous,aghion2020infinite}. The relation in Eq. \eqref{LAndMID}, if observed in data, can in-fact be used to indicate that the underlying process has an infinite-invariant density in this regime, and it was also observed in the Pommeau-Manneville map \cite{meyer2017infinite}. If \eq{\alpha>0} and \eq{\beta\neq0}, or equivalently  \eq{L>1/2} and $M\neq$[Eq. \eqref{LAndMID}], the limit shape of the increment PDF is given by an infinite-covariant density, see e.g.,  \cite{kessler2010infinite,lutz2013beyond,rebenshtok2014non,holz2015infinite,aghion2017large,wang2019transport,aghion2018asymptotic}.  
 
 Note that, in both the invariant and the covariant case, and also in the case when the mean-absolute and mean-squared increments are non-divergent, but they correspond to  different scaling regimes of the PDF, a Noah effect cannot appear without  a Moses effect.
 The different cases for $M,L$ and $\alpha,\beta$ are summarized in Table~\ref{tablelAlphaBeta}.  
 
\begin{table}[t] 
\footnotesize
\centering 
\begin{tabular}{c c c c } 
\hline\hline 
$\alpha$&$\beta$&$L,M$&\eq{\lim_{t\rightarrow\infty}t^{\alpha+\beta}P(\MeanV/t^\beta)}\\
[0.5ex] %
\hline 
$0$&$0$&$\frac{1}{2},\frac{1}{2}$&steady-state\\
$0$&$\beta\neq0$&$\frac{1}{2},M>\frac{1}{2}$&normalized scaling limit\\
$\alpha>0$&$0$&$L>\frac{1}{2},M<\frac{1}{2}$&infinite-invariant density\\
$\alpha>0$&$\beta\neq0$&$L>\frac{1}{2},(all)$&infinite-covariant density\\ 
[0.5ex]
\hline\hline 
\end{tabular} 
\caption{\footnotesize{Summary of the different scaling limit of $P_t(\MeanV)$, that can be found from the Moses $M$ and Latent $L$ exponents, via $\alpha,\beta$ Eqs. (\ref{ICD},\ref{ALphaBetaDefinitions}), if both \eq{\langle|\MeanV|\rangle} and $\langle \MeanV^2\rangle$ correspond to the same scaling regimes of the PDF. Note that $\alpha,\beta$ set the restrictions for $M,L$ in the various regimes, not the other way around. 
}}
\label{tablelAlphaBeta} 
\end{table} 

\section{The L\'evy walk model}
\label{SecModel} 

As mentioned in the introduction, in this work we analyse a two-state L\'evy walk model. Particularly, here, we consider a continuous range of IID random step velocities, whose distribution is \eq{\phi(V)}. In addition, we assume a nonlinear coupling between the \eq{i}th step duration and the step velocity, namely 
\begin{equation}
V_i= \pm\tilde{c}_1\tau_i^{\nu-1}, 
    \label{NonlinearDurationVelocityCoupling}
\end{equation} 
where 
\begin{equation} 
\nu>0. 
\label{NuRegime} 
\end{equation}
The sign of the step velocity is randomly chosen to be positive or negative with equal probability (the motion is unbiased). The constant \eq{\tilde{c}_1} has units of \eq{distance/(time)^\nu}, but throughout this manuscript we set \eq{\tilde{c}_1=1} for convenience.
 Eq. \eqref{NonlinearDurationVelocityCoupling} means that \EQ{\phi(V)=\frac{1}{2}\int_0^\infty\Intd \tau g(\tau)\left[\delta(V-\tau^{\nu-1})+\delta(V+\tau^{\nu-1})\right].}{StepVelocityPDF} Below, in our numerical simulations, we will use a specific example  where the IID random step durations are obtained from the distribution 
\begin{equation}
g(\tau)=\gamma\tau_0^\gamma\tau^{-1-\gamma}\Theta(\tau\geq\tau_0),  
    \label{MLDistributionOfTaus}
\end{equation}  
though our results are more general (see the discussion, Sec. \ref{Discussion}). Here, \eq{\tau_0>0} can be as small as we wish, and \eq{\Theta(\cdot)\equiv1} when the condition inside the brackets is satisfied and zero otherwise. 
For any \eq{g(\tau)} in Eq. \eqref{gTauDefinition}, from Eqs. (\ref{NonlinearDurationVelocityCoupling},\ref{StepVelocityPDF}), when \eq{|V|<1} one finds that   \eq{\phi(V)\sim \frac{c}{2(1-\nu)|\Gamma(-\gamma)|}|V|^{-1-\gamma/(\nu-1)}}. For our example, from Eq. \eqref{MLDistributionOfTaus} it follows that the step velocity distribution in the first \eq{N-1} complete steps, when $\nu<1$, is  
\begin{equation} 
\phi(V)=\frac{\gamma\tau_0^\gamma}{2(1-\nu)}{|V|^{-\frac{\gamma}{\nu-1}-1}}\Theta(|V|\leq\tau_0^{\nu-1}),   
    \label{StepVelocityPDF1}
\end{equation} 
and it has a similar shape but with~$\Theta(|V|\geq\tau_0^{\nu-1})$ replacing the original one when $\nu>1$, hence \eq{c=\gamma \tau_0^\gamma|\Gamma(-\gamma)|}. In this manuscript we  focus on the parameter regime 
\begin{equation} 
0<\gamma<1, 
\label{ParameterRegime} 
\end{equation} 
where \eq{\langle\tau\rangle} is divergent. In various models of non-linearly coupled L\'evy walk, some of them are summarized in the review  \cite{LevyWalks}, it was shown that in addition to the various scaling exponents, the statistical properties of the process depend strongly on the  treatment given to the last, incomplete, step in the sequence. We choose to correspond with the model studied in \cite{albers2018exact,akimoto2019infinite,bothe2019mean}, where \eq{V_N} is determined from the time interval straddling \eq{t} \cite{wang2018renewal}. With this choice, all the velocities \eq{V_i}, with \eq{i=1..N} are IID, though the duration of the last step is given by Eq. \eqref{TotalTime}.  As usual, the displacement at each step (complete and incomplete) is the linear product of the step velocity and its duration.

{\em Instantaneous velocity PDF.}
Akimoto et al. \cite{akimoto2019infinite}, studied the instantaneous velocity PDF \eq{P_t(v)} of the L\'evy walker in the process described above, at time \eq{t\gg1} and the regime where $0<\nu<1$. We can apply their results to our analysis, since in this model we can associate $\MeanV$ and $v$ via $v=\lim_{\Delta\rightarrow0}\MeanV$, see Sec. \ref{SecMAndL}. The following analytic results are brought from that referenced paper. At long but finite times, $P_t(v)$ assumes different shapes in two separate ranges of \eq{v}:  Let \eq{v_c= t^{\nu-1}}, then \cite{akimoto2019infinite}
\begin{equation} 
P_t(v)\approx \begin{cases} \frac{t^\gamma}{2(1-\nu)|\Gamma(-\gamma)|\Gamma(1+\gamma)}|v|^{-1-\gamma/(\nu-1)}, & |v|/v_c\leq 1\vspace{5pt}\\ 
\frac{1-[1-(v/v_c)^{1/(\nu-1)}]^\gamma}{c \Gamma (\gamma+1) }t^\gamma\phi(v), & |v|/v_c>1.  
\end{cases} 
\label{VelocityPDFSmallV}
\end{equation} 
Due to the asymptotic shape of \eq{\phi(v)}, when \eq{v} itself is smaller than unity (regardless of \eq{t}),  $P_t(v/t^{\nu-1})$ corresponds in this regime to the scaling function \eq{\sim t^{(\nu-1)}\rho(\tilde{v})}, where \eq{\tilde{v}= v/t^{\nu-1}}  and 
\begin{equation} 
\rho(\tilde{v})\approx \begin{cases} \frac{1}{2(1-\nu)|\Gamma(-\gamma)|\Gamma(1+\gamma)}|\tilde{v}|^{-1-\gamma/(\nu-1)}, & |\tilde{v}|\leq 1\vspace{5pt}\\ 
\frac{1-[1-(\tilde{v})^{1/(\nu-1)}]^\gamma}{2(1-\nu)|\Gamma(-\gamma)| \Gamma (\gamma+1) }|\tilde{v}|^{-1-\gamma/(\nu-1)}, & |\tilde{v}|>1.  
\end{cases} 
\label{WzBetaSmallV}
\end{equation} 
The scaling function $\rho(\tilde{v})$ is normalized to unity. 
On the other hand, at long times $P_t(v)$ has a second scaling shape valid in the region $v>v_c$, since in the limit \eq{t\rightarrow\infty} the support of the region $v/v_c<1$ in Eq. \eqref{VelocityPDFSmallV} goes to zero, and at  \eq{v/v_c\gg1}, we can expand \eq{[1-(v/v_c)^{1/(\nu-1)}]^\gamma} as a Taylor series for the small parameter \eq{(v/v_c)^{1/(\nu-1)}}. This yields, to leading order in time, 
$
P_t(v)\approx \phi(v)|v|^{\frac{1}{\nu-1}}\frac{t^{\gamma-1}}{c\Gamma(\gamma)}. $  
Which means that  asymptotically   \cite{akimoto2019infinite}, 
\begin{equation} 
\lim_{t\rightarrow\infty}t^{1-\gamma}P_t(v)\rightarrow \mathcal{I}(v),\Hquad\mbox{where}\Hquad \mathcal{I}(v)\equiv\phi(v)|v|^{\frac{1}{\nu-1}}\frac{1}{c\Gamma(\gamma)}, 
\label{VelocityID} 
\end{equation} 
and $\phi(v)$ is in Eq. \eqref{StepVelocityPDF1}.
The time-invariant asymptotic limit given by \eq{\mathcal{I}(v)} in Eq. \eqref{VelocityID} is  non-integrable around \eq{v=0}, hence it is non-normalizable:   \eq{\int_{-\infty}^\infty\mathcal{I}(v)\Intd v
\rightarrow\infty.} As such, this function  is the infinite-invariant density of the process \cite{akimoto2019infinite}. Note that when $\nu>1$, the two regimes of the PDF, Eq. \eqref{VelocityPDFSmallV} simply switch places, but their functional shape remains the same. 



 
\begin{figure}
\centering
\includegraphics[width=0.38\textwidth]{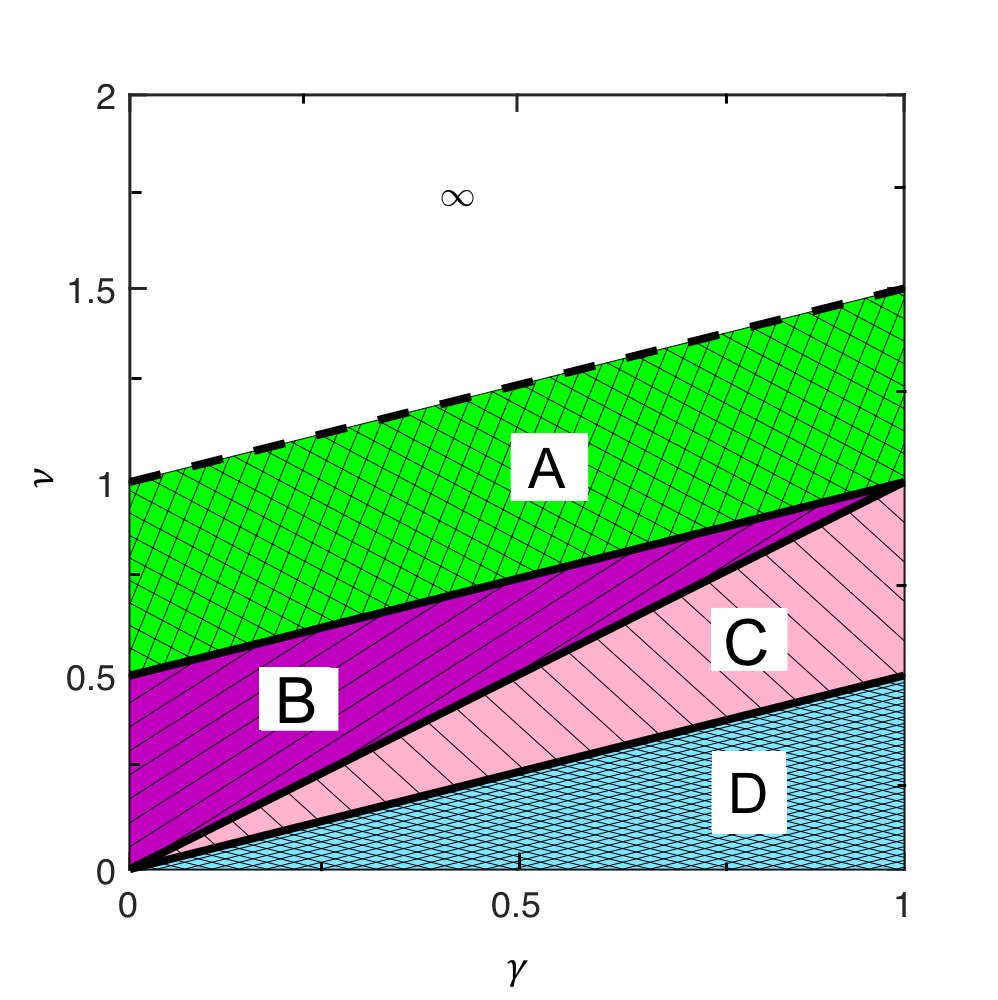}
 \caption{\footnotesize{Phase diagram of the scaling exponents describing the decomposition of the anomalous diffusion. The three solid lines, separating regions $A$-$B$, $B$-$C$ and $C$-$D$  
 are respectively:
 \eq{\nu=\gamma/2+1/2}, \eq{\nu=\gamma} and \eq{\nu=\gamma/2}. The dashed-line is \eq{\nu=\gamma/2+1}. The results for the three-effect decomposition in the various regimes are discussed in Sec. \ref{SecMainResults}.  Region A: $H = \nu$, $J = 1$, $L = 1/2$, $M = \nu-1/2$  (``maximal" Joseph effect, namely the autocorrelation function Eq. \eqref{JosephDefinitionContinousTime} does not decay at large values of $\Delta$, no Noah, $P_t(\MeanV)$ has a normalized scaling shape corresponding to $M$). Region B: $H = \nu$, $J = (1+2\nu-\gamma)/2$, $L = 1-\nu+\gamma/2$, $M = \nu-1/2$ (onset of a Noah effect). Region C: $H = \nu$, $J = (1+2\nu-\gamma)/2$, $L = 1- \gamma/2$, $M = \gamma-1/2$ ($P_t(\MeanV)\rightarrow\mbox{infinite-invariant density}$). Region D: $H = \gamma/2$, $J = 1/2$, $L = 1- \gamma/2$, $M = \gamma-1/2$ (infinite-invariant density, no Joseph effect). In the $``\infty''$ regime, $H\rightarrow\infty$, and $M,L,J$ are not well defined.
 }}
 \label{fig2:QgphaseAllTogether}
\end{figure}
 
 \section{Summary of our main results} 
 \label{SecMainResults}
This summary brings the main results of our analysis of  L\'evy walk trajectories generated by the process described in Sec. \ref{SecModel}, and the detailed derivations appear below. For further discussion about the generality of the three-effect decomposition, also see below. 
In our simulations, we generated an ensemble of $10^8$ realizations of the process $x(t)$ for different values of $\gamma$ and $\nu$, and observed the increments $\delta x_{j}$ of the paths at different times ranging from $t=10^4$ to $10^8$. We then measured the ensemble averages of $|\delta x_{j}|$,  $\delta x_{j}^{2}$ (namely, we used $\MeanV$ with observation windows of duration $\Delta=1$), as well as $x^2$, to calculate the values of $M$, $L$ and $H$ respectively. To obtain the value of the exponent $J$, we used a method based on the time-averaged MSD $\delta^2$, as explained in detail in Sec. \ref{JosephExponent} and Appen. \ref{Appenctamsd}. The results of this method correspond to those of a direct measurement of the correlation function, but it is numerically more convenient (see Appen. \ref{Appenctamsd}).

{\em What the data analysis says:} \textbf{Without} relying on prior knowledge about the underlying process, we found that in the range defined by Eqs. (\ref{NuRegime},\ref{ParameterRegime}),  the L\'evy walk data exhibits five separate dynamical phases. These phases are summed-up below and in Fig. \ref{fig2:QgphaseAllTogether}. The summation formula, Eq. \eqref{hjlm} is confirmed in all but the ``$\infty$" regime. 
\begin{itemize}
    \item In regime A, when $\gamma/2+1/2<\nu<\gamma/2+1$: $H = \nu$, $J = 1$, $L = 1/2$, $M = \nu-1/2$. Here, the auto-correlation function does not decay with \eq{\tilde{\Delta}}, in Eq. \eqref{JosephDefinitionContinousTime}, namely the increments are essentially completely correlated. In this situation, we say that the Joseph effect is maximal, since by definition $J$ can never be bigger than its value here. There is no freedom left in the increment distribution for any Noah effect to be present. There can be, however, a Moses effect as the increment distribution does ``age" with time.
    The existence of a Moses effect without a Noah effect means that in this regime we expect a single scaling function in the form of $t^{\nu-1}P_t(\MeanV/t^{\nu-1})$ to describe the regime of the PDF which gives rise to the first and second moments of $|\MeanV|$ (which is therefore no-fat tailed). Our numerics show that this regime extends also to the range $1<\nu<\gamma/2+1$ (and $\gamma<1$). 
    
    \item In regime B, $\gamma<\nu<\gamma/2+1/2$: $H = \nu$, $J = (1+2\nu-\gamma)/2$, $L = 1-\nu+\gamma/2$, $M = \nu-1/2$. In this regime all the three effects contribute to the anomalous diffusion. 
    Here, the Joseph effect is present, but is not maximal, as the auto-correlation function decays as a power-law function of \eq{\tilde{\Delta}}. 
    This allows for a Noah effect to be present too. Here, the Noah effect means that the scaling shape at the bulk of \eq{P_t(\MeanV)} is fat-tailed, in the sense that its second moment is divergent.  But the mean of $|\MeanV|$ remains unchanged from regime A, so it is expected to still be given by the same scaling regime of the increment PDF as before, namely $\langle|\MeanV|\rangle$ and $\langle\MeanV^2\rangle$ correspond to different regimes of $P_t(\MeanV)$. Accordingly our numerical analysis shows that Eq. \eqref{ICD} is not valid in this case. The Moses effect occurs here in a similar way as it does in regime A, namely also in this regime, the increment PDF is not time-invariant.

\item In regime C,  $\gamma/2<\nu<\gamma$: $H = \nu$, $J = (1+2\nu-\gamma)/2$, $L = 1- \gamma/2$, $M = \gamma-1/2$. Still, all three effects contribute to the anomalous diffusion. 
   Here, just as in regime B, the Joseph effect is present, but is not maximal. 
In this regime, the Moses and Noah effects are coupled, with the Moses and the Latent exponents obeying Eq. \eqref{LAndMID}. This suggests that the large fluctuations of the system are described by an infinite-invariant density, Eq. \eqref{ICD} with $\alpha=1-\gamma, \beta=0$. 

\item In regime D,   \eq{\nu<\gamma/2}: $H = \gamma/2$, $J = 1/2$, $L = 1- \gamma/2$, $M = \gamma-1/2$. Here, $M,L$ remain coupled as in region C. Hence we expect the same infinite-invariant density to be valid in this regime too. Interestingly, now there are no long-range increment correlations and, thus, there is no Joseph effect. At this stage anomalous diffusion occurs due to the non-stationarity of $P_t(\MeanV)$ and the fat tails of the scaling-shape describing this PDF at the bulk.  

\item When $\nu>\gamma/2+1$, the MSD is divergent. The scaling relations in Eqs. (\ref{NoahDefinitionContinousTime}-\ref{JosephDefinitionContinousTime})  don't hold, and in this regime the decomposition is not valid. We call this the ``$\infty$" regime. See Appen.~\ref{AppendixInfinityRegime}.

\end{itemize} 

{\em What we know from the model, in comparison with the data analysis:}  When $\gamma,\nu<1$, Eq. \eqref{WzBetaSmallV} and Eq. \eqref{VelocityID} describe two different ways to obtain a time-invariant scaling-shape of the instantaneous velocity PDF $P_t(v)$, the first is valid for small $v$ and the second for large. We can associate this velocity PDF with the distribution of the increment velocity $\MeanV$ (see Sec. \ref{SecMAndL}). As expected from the numerics, the analytic results presented in Sec. \ref{SecMAndL} show that the bulk function and the infinite-invariant density describe the shape of the increment PDF in regimes $A$ and $C,D$ respectively, in the range of $v$ which is responsible for the various moments. In regime B, $\langle|v|\rangle,\langle v^2\rangle$, (hence $\langle|\MeanV|\rangle,\langle \MeanV^2\rangle$) are obtained separately from the two scaling regimes. The fact that the Joseph effect, studied in Sec. \ref{JosephExponent}, is ``maximal'' in regime A, matches to the fact that the bulk limit-function describing the PDF is thin-tailed, from the same reason that in regime D it is ``minimal": if the increments are long- (short-) ranged correlated, their size is more (less) predictable from the first step. Therefore, large fluctuations are less (more) possible. 

In regime A, when $\nu>1$, it is easy to show that one can find similar results for $\langle|\MeanV|\rangle$ and $\langle\MeanV^2\rangle$ as in the case when $\nu<1$ since as mentioned, the shape of $P_t(\MeanV)$ is similar to Eq. \eqref{VelocityPDFSmallV}, but with the two regimes for $v\leq1$ and $v>1$ switching roles. In addition, here $\tau_c^{\nu-1}$ becomes a lower, instead of an upper cutoff for the step velocity PDF in Eq. \eqref{StepVelocityPDF1}. The divergence of the MSD in the ``$\infty$'' regime, was shown analytically in \cite{albers2018exact,bothe2019mean}, further details in Sec. \ref{Section5} and Appen. \ref{AppendixInfinityRegime}.

\section{\eq{M}\& \eq{L}, and how we obtained them} 
 \label{SecMAndL}


As explained in Sec. \ref{SecThreeExps}, in order to obtain \eq{M} and \eq{L}, we need to examine the temporal behavior of the ensemble-time averages  \eq{\left\langle\overline{|\MeanV|}\right\rangle} and \eq{\left\langle\overline{\MeanV^2}\right\rangle}, where \eq{\MeanV} is the mean velocity obtained at increments \eq{\delta x}, whose duration \eq{\Delta} is defined independently from step duration of the underlying L\'evy walk (namely \eq{\Delta\neq\tau}). Choosing  \eq{\Delta\ll1}, the mean velocity \eq{\MeanV} can be exchanged with the instantaneous velocity \eq{v} of the random walker at various points in time, and then we can replace \eq{\MeanV\leftrightarrow v} in Eqs. (\ref{MosesDefinitionContinousTime}-\ref{NoahDefinitionContinousTime}). Accordingly, this means that we can obtain the exponents of the time series from $\left\langle\overline{|v|}\right\rangle$ and $\left\langle\overline{|v|^2}\right\rangle$, where we now use the following definition for the time average of an observable \eq{f}: $\overline{f}=(1/t)\int_0^t f(t')\Intd t'$. We note that here one should use a bit of care, since during an increment of duration \eq{\Delta}, the particle might have ended one step of the underlying random walk, and started another, and in this interval of the motion the mean velocity is different from the instantaneous value before/ after the transition. However we assume that if \eq{\Delta} is small enough, the effect of these occurrences is negligible in the context of the results in this manuscript. This is also confirmed by our numerics. 

\begin{figure*}
\centering
\includegraphics[width=0.85\textwidth]{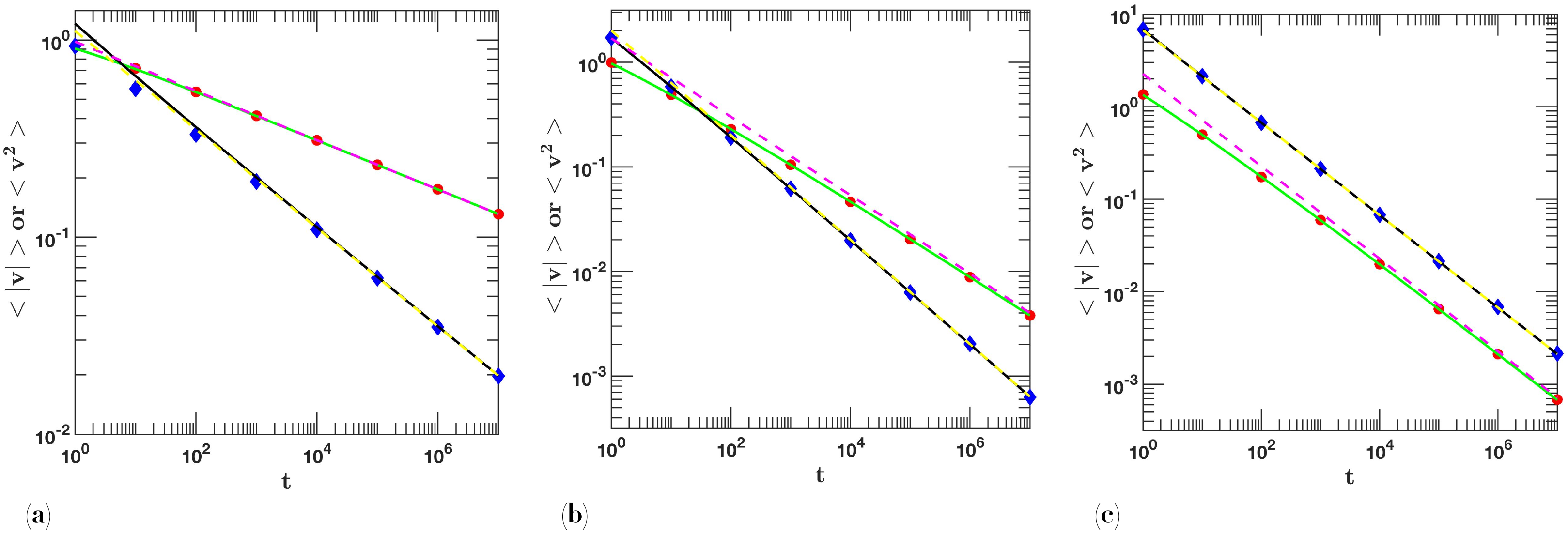}
\caption{\footnotesize{Log-log plots for the averages of $|v|$ and $v^2$, as function of  time. Red dots and the blue diamonds represent the values of $<|v|>$ and $<v^2>$, obtained from simulated data for different values of $t$, respectively. The solid green and black lines correspond to Eq. \eqref{EA|V|MiddleRegime} and Eq. \eqref{MeansqrvNonIntegrableRegime} respectively. The yellow and the magenta dashed lines represent the leading order terms in these equations, in the long time limit. (a) gives the result in the integrable regime, Sec. \ref{IntegrableRegime}, with $\gamma = 0.5$ and $\nu = 0.875$, (b) the middle regime, Sec. \ref{MiddleRegime}, with $\gamma = 0.5$ and $\nu = 0.625$, and (c) the non-integrable regime, Sec. \ref{NonintegrableRegime}, with $\gamma = 0.5$ and $\nu = 0.375$. The simulation results were generated with $10^8$ realizations and $\tau_c = 0.01$.}}
 \label{figEnsembleMoments}
\end{figure*}

\subsection{Three regimes for \eq{M} and \eq{L}}

One can obtain the long-time asymptotic behavior of the ensemble mean of any symmetric observable \eq{\mathcal{O}(v)} in the system, as follows: 
\begin{align}
    \langle \mathcal{O}(v)\rangle&= 2\int_0^\infty \mathcal{O}(v)P_t(v)\Intd v\nonumber\\ 
    &=2\int_0^{v_c}\mathcal{O}(v)P_t(v)\Intd v+
    2\int_{v_c}^{\infty}\mathcal{O}(v)P_t(v)\Intd v. 
    \label{EnsembleMeanOfFv}
\end{align} 
Given Eqs. (\ref{StepVelocityPDF1},\ref{VelocityPDFSmallV}), for the mean of $|v|$, we get 
\small 
\begin{align}
 \langle |v|\rangle&\approx 2\int_0^{v_c}\frac{1}{2(1-\nu)|\Gamma(-\gamma)|\Gamma(1+\gamma)}t^\gamma|v|^{-\frac{\gamma}{\nu-1}}\Intd v\nonumber\\ 
 &+2\int_{v_c}^{{\tau_c}^{\nu-1}}\frac{1-[1-(v/v_c)^{1/(\nu-1)}]^\gamma}{2(1-\nu)|\Gamma(-\gamma)| \Gamma (\gamma+1) }t^\gamma{|v|^{-\frac{\gamma}{\nu-1}}}\Intd v   
\nonumber\\ 
     &\approx-\frac{  \Gamma (-\gamma+\nu-1)t^{\nu-1}}{|\Gamma(-\gamma)|   \Gamma ({\nu})}+\frac{  {\tau_c}^{\nu-\gamma}t^{\gamma-1}}{|\Gamma(-\gamma)|\Gamma(\gamma)  (\gamma-\nu)}. 
\label{EA|V|MiddleRegime} 
\end{align} 
\normalsize  
 Similarly, for the mean of $v^2$, we get 
\small 
\begin{align}
    \langle v^2\rangle&\approx 2\int_0^{v_c}\frac{t^\gamma}{2(1-\nu)|\Gamma(-\gamma)|\Gamma(1+\gamma)}|v|^{1-\frac{\gamma}{\nu-1}}\Intd v\nonumber\\ 
 &+2\int_{v_c}^{{\tau_c}^{\nu-1}}\frac{1-[1-(v/v_c)^{1/(\nu-1)}]^\gamma}{2(1-\nu)|\Gamma(-\gamma)| \Gamma (\gamma+1) }t^\gamma{|v|^{1-\frac{\gamma}{\nu-1}}}\Intd v   
\nonumber\\ 
     &\approx-\frac{ \Gamma (-\gamma+2 \nu-2)t^{2 \nu-2}}{\Gamma (2 \nu-1) \left| \Gamma (-\gamma)\right| }+\frac{{\tau_c}^{-\gamma+2 \nu-1}t^{\gamma-1}}{(\gamma-2 \nu+1) \Gamma (\gamma) \left| \Gamma (-\gamma)\right| }. 
    \label{MeansqrvNonIntegrableRegime}
\end{align} 
\normalsize
To determine the leading behavior of these two means in the long time limit, note that Eqs. (\ref{WzBetaSmallV},\ref{VelocityID}) create a distinction between two different cases, depending on whether $\OO({v})=|v|$ or $v^2$ is integrable with respect to $\rho(v)$, Eq. \eqref{WzBetaSmallV}, or it is integrable with respect to the infinite-invariant density \eq{\mathcal{I}(v)}, Eq. \eqref{VelocityID}. In the first case, the leading order is obtained by first changing variables: $v/t^{\nu-1}\rightarrow\tilde{v}$ $\langle\OO(v)\rangle=2t^{q(\nu-1)}\int_0^{1/t^{\nu-1}}\OO(\tilde{v})\rho(\tilde{v})\Intd \tilde{v}+2t^{q(\nu-1)}\int_{1/t^{\nu-1}}^\infty\OO(\tilde{v})P_t(\tilde{v} t^{\nu-1})\Intd \tilde{v}$, and then in the range $t\gg1$ the first term is $\approx 2t^{q(\nu-1)}\int_0^{\infty}\OO(\tilde{v})\rho(\tilde{v})\Intd \tilde{v}$ and the second term approaches zero since its support vanishes. So, in this case  
\small
\begin{equation} 
    \langle \mathcal{O}(v)\rangle\approx 
    2t^{q(\nu-1)}\int_0^{\infty}\OO(\tilde{v})\rho(\tilde{v})\Intd \tilde{v}.
    \label{EnsembleMeanOfFvWzBeta} 
\end{equation} 
\normalsize 
In the second case, when \eq{\langle \mathcal{O}(v)\rangle_{\mathbb{I}}\equiv\int_0^\infty \mathcal{O}(v)\mathcal{I}(v)\Intd v<\infty}, \eq{\mathcal{O}(v)} is integrable with respect to the infinite-invariant density, the contribution to its mean from the region \eq{v<v_c} can be neglected in Eq. \eqref{EnsembleMeanOfFv} in the limit \eq{t\rightarrow\infty}, to leading order, hence using Eq. \eqref{VelocityID} we get 
\begin{equation}
    \langle\mathcal{O}(v)\rangle\approx 
    2\int_{v_c\rightarrow0}^\infty \mathcal{O}(v)P_t(v)\Intd v\underbrace{\rightarrow}_{t\rightarrow\infty} 2t^{\gamma-1}\langle \mathcal{O}(v)\rangle_{\mathbb{I}}.
\label{Eq20ID}
\end{equation}
Notice that in this case, the temporal scaling of  \eq{\langle\OO(v)\rangle} is similar for all the integrable observables, since it is determined only by the scaling of the infinite-density. 
For \eq{\mathcal{O}(v)=\langle |v|\rangle} and \eq{\langle v^2\rangle} together, there are three regimes of behavior, included within the range \eq{\gamma,\nu<1}: The integrable regime, where both $\langle|v|\rangle$ and $\langle v^2\rangle$ are integrable with respect to \eq{\rho(v)}; The middle regime, where only the mean-absolute velocity is integrable;  And the non-integrable regime, where neither observable is integrable (details below). 
Figs. \ref{figEnsembleMoments}a-c display simulation results for the temporal behaviour of $\langle|v|\rangle$ and $\langle v^2\rangle$ in the integrable, the middle and the non-integrable regimes, respectively. The simulations match perfectly at long times with the exact expressions in Eqs. (\ref{EA|V|MiddleRegime},\ref{MeansqrvNonIntegrableRegime}), which denote both the leading order behavior of $\langle|v|\rangle$ and $\langle v^2\rangle$ in time, and the next-to-leading order. They also confirm the approach to the leading order asymptotic results, though this approach is slow. The results for the exponents $M$ and $L$ in the various regimes are shown in the lower two panels of Fig. \ref{FigOfFourPhaseDiagrams}. In Fig. \ref{FigICD}, we use the results for these exponents in the three regimes, in order to seek for a time invariant asymptotic shape of $P_t(\MeanV)$, based on Eqs. (\ref{ICD},\ref{ALphaBetaDefinitions}).


\subsection{The integrable regime, \eq{1/2+\gamma/2<\nu<1}} 
\label{IntegrableRegime} 

In this regime, the leading behavior in time of \eq{\langle|v|\rangle} and $\langle v^2\rangle$ is given by the $\sim t^{\nu-1}$ and $\sim t^{2\nu-2}$ terms in Eq. \eqref{EA|V|MiddleRegime} and Eq. \eqref{MeansqrvNonIntegrableRegime}, respectively. The second term in both equations gives the next-to-leading order behavior. This result agrees with the calculation based on Eq. \eqref{EnsembleMeanOfFvWzBeta}. Similar to the argument in Eq. \eqref{EATAVsEA}, the ensemble-time averages \eq{\langle\overline{|v|}\rangle\propto t^{\nu-1}} and \eq{\langle\overline{v^2}\rangle\propto t^{2\nu-2}}, like their corresponding ensemble averages, and since we associate \eq{v} with \eq{\MeanV}, we now obtain the Latent and Moses exponents using Eqs. (\ref{MosesDefinitionContinousTime},\ref{NoahDefinitionContinousTime}):  
\begin{equation} 
M=\nu-\frac{1}{2},\qquad\mbox{and}\qquad L=\frac{1}{2}. 
    \label{MosesNoahNonIntegrableRegime}
\end{equation} 

Since here both means are obtained from the same scaling limit of $P_t(v)$, we can now associate \eq{\rho(\tilde{v})} in this regime with $W(z_\beta)$, Eq. \eqref{ICD}, and here \eq{z_\beta=\tilde{v}=v/t^{\nu-1}}, so $\beta=\nu-1$ and $\alpha=0$, in agreement with Eqs. (\ref{ALphaBetaDefinitions},\ref{MosesNoahNonIntegrableRegime}).  
Fig. \ref{FigICD}a displays the convergence of simulation results of $P_t(v)$ at increasing times, rescaled according to Eq. \eqref{ICD}, as function of $z_\beta$, to the scaling limit Eq. \eqref{WzBetaSmallV}. 
Note that the Moses effect originates from the diverging mean duration of the L\'evy walk steps, namely because \eq{\langle\tau\rangle\rightarrow\infty} in \eq{g(\tau)}, Eq. \eqref{MLDistributionOfTaus}, which leads to statistical aging \cite{godreche2001statistics}. 

\subsection{Middle regime, \eq{\gamma<\nu<1/2+\gamma/2}}
\label{MiddleRegime}
In this regime, \eq{v^2} is no longer integrable with respect to the scaling function $\rho(v)$. $|v|$, however, still is.  Therefore, the leading-order behavior of $\langle|v|\rangle$ and $\langle\overline{|v|}\rangle$, remains proportional to $\sim t^{\nu-1}$, similar to  the previous, integrable region. However since $\langle v^2\rangle$ is now integrable with respect to the infinite-density $\mathcal{I}(v)$ instead of $\rho(v)$, its leading behavior is now obtained from Eq. \eqref{Eq20ID}. The result is equal to the term $\propto t^{\gamma-1}$ in Eq. \eqref{MeansqrvNonIntegrableRegime} (and the second term there is now the next-to-leading order behaviour). Therefore, also $\langle\overline{v^2}\rangle\sim t^{\gamma-1}$. Note that in this regime we can obtain the time average of $v^2(t)$ also using arguments based on infinite-ergodic theory \cite{akimoto2019infinite}.  
 Using Eqs. (\ref{MosesDefinitionContinousTime},\ref{NoahDefinitionContinousTime}), this yields 
\begin{equation} 
M=\nu-\frac{1}{2},\qquad\mbox{and}\qquad L=\frac{\gamma-2\nu+2}{2}. 
    \label{MosesNoahMiddleRegime}
\end{equation} 

This regime continuously extends the one introduced in Eq. (\ref{MosesNoahNonIntegrableRegime}). The First moment can still be described by the scaling shape of the PDF at the bulk. However, since the second moment of this PDF diverges with respect to $\rho(v)$, here we see for the first time the emergence of a Noah effect, in addition to Moses. 
Since the mean of $|v|$ and $v^2$ are obtained from two different scaling regimes of $P_t(v)$, Eqs. (\ref{ICD}-\ref{ICD1}) are not valid, and $\alpha$ and $\beta$ are not defined. Fig. \ref{FigICD}b shows that, if we did not know the model, and try to obtain $\alpha,\beta$ from Eq. \eqref{ALphaBetaDefinitions} in this regime from the data, we would find $\alpha=\gamma-2\nu+1,\beta=\gamma-\nu$, but with this rescaling, $P_t(v)$ does not converge to a time-invariant shape. 
\begin{figure*}
\centering
(a)\includegraphics[width=0.32\textwidth]{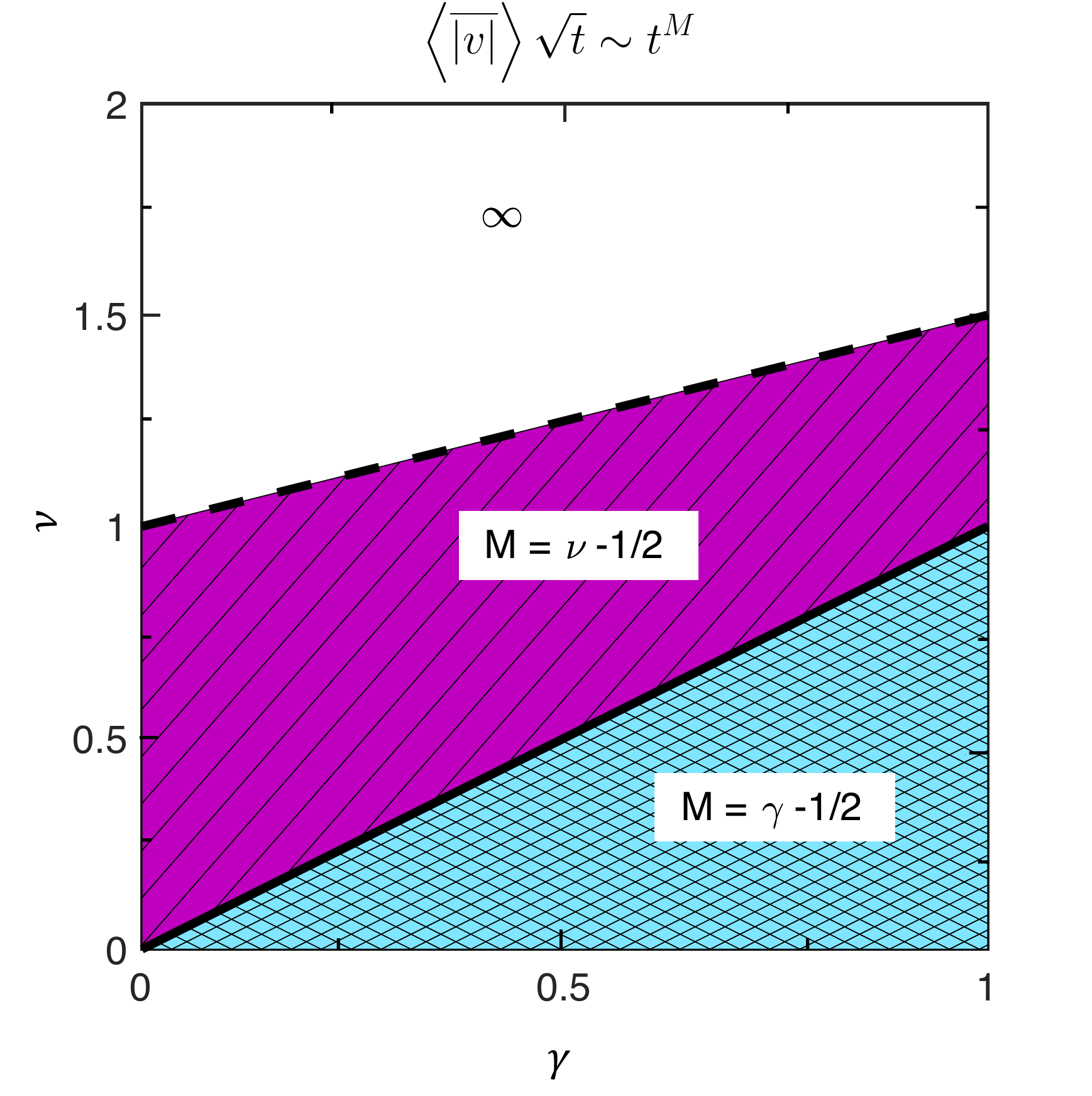}
(b)\includegraphics[width=0.32\textwidth]{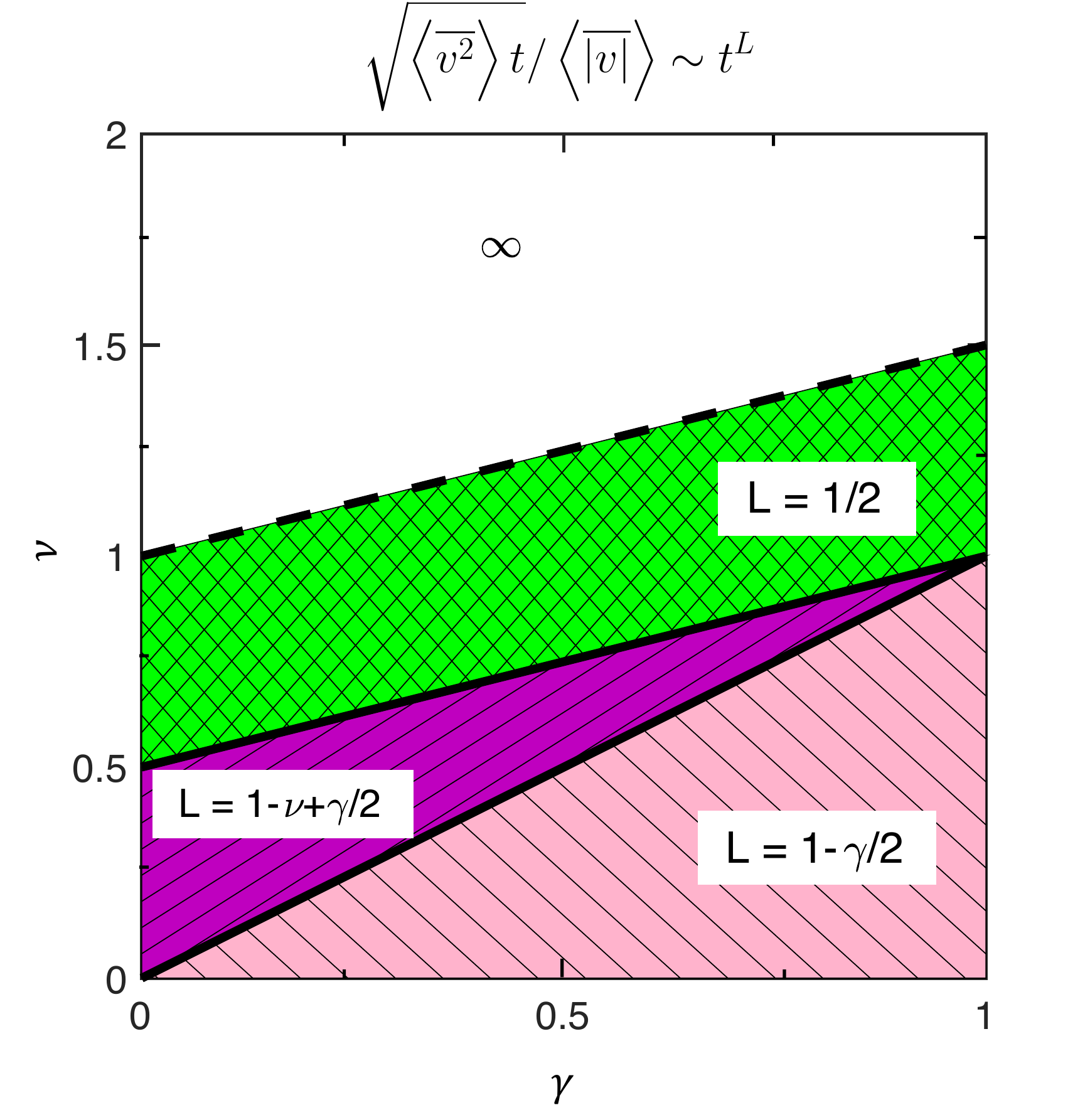}
(c)\includegraphics[width=0.32\textwidth]{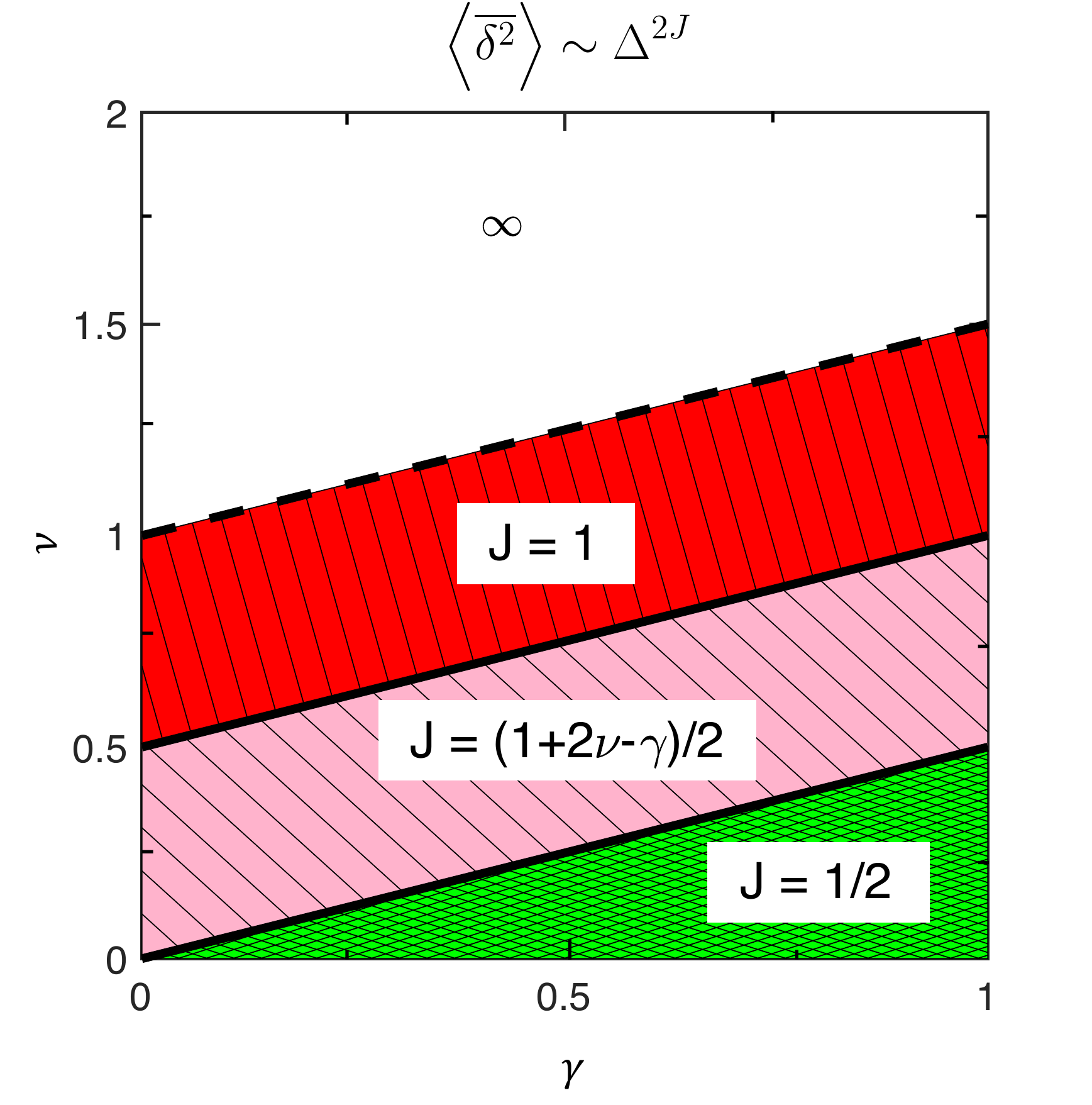}
(d)\includegraphics[width=0.32\textwidth]{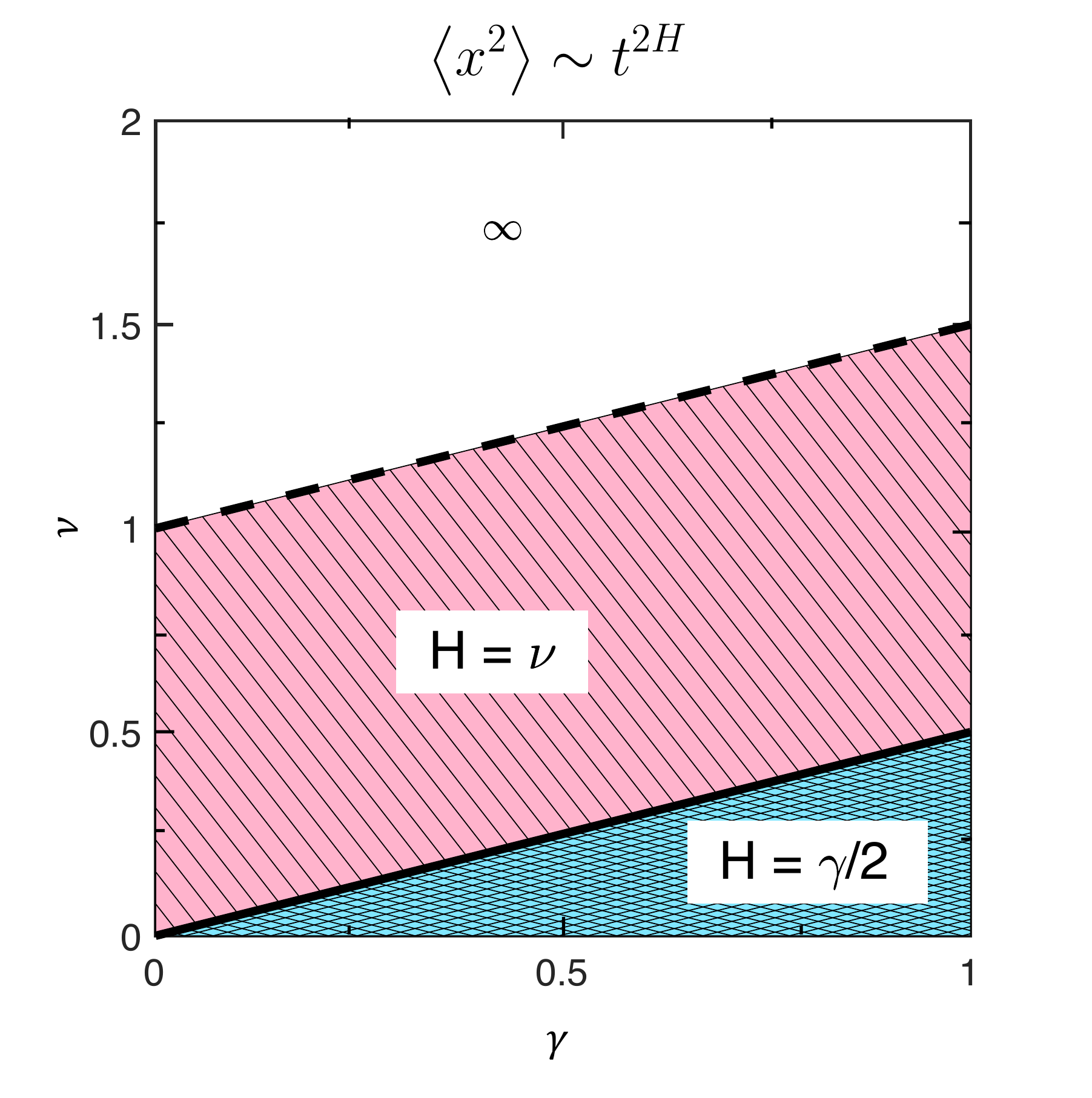}
 \caption {\footnotesize{Phase diagrams of the scaling exponents describing the decomposition of the anomalous diffusion of L\'evy walks into three constitutive effects, and their various magnitudes. (a) gives the Moses exponent $M$ that quantifies the Moses effect, (b) gives the Latent exponent $L$ that quantifies the Noah effect, (c) gives the Joseph exponent $J$ that quantifies the Joseph effect, and (d) gives the Hurst exponent $H$.  These results were obtained with for $\eta$ = 1 (see Sec. \ref{Section5}). In the ``$\infty$" regime, the Hurst exponent is divergent and the other exponents are not well defined, see Sec. \ref{Section5} and Appen. \ref{AppendixInfinityRegime}}}
 \label{FigOfFourPhaseDiagrams}
\end{figure*} 
\subsection{The non-integrable regime, \eq{\nu<\gamma}}
\label{NonintegrableRegime}
In this regime neither the first, nor the second moment of $|v|$ are integrable with respect to $\rho(v)$, Eq. \eqref{WzBetaSmallV}. Instead, both the mean velocity, and the mean squared velocity are integrable with respect to the infinite-density. Here,  using Eqs. (\ref{StepVelocityPDF1},\ref{VelocityID},\ref{Eq20ID}) we get 
$\langle |v|\rangle,\langle \overline{|v|}\rangle\propto t^{\gamma-1}$, as well as $\langle v^2\rangle,\langle \overline{v^2}\rangle\propto t^{\gamma-1}$, so from Eqs. (\ref{MosesDefinitionContinousTime},\ref{NoahDefinitionContinousTime}), we find 
\begin{equation} 
M=\gamma-\frac{1}{2},\qquad\mbox{and}\qquad L=1-\frac{\gamma}{2}. 
    \label{MosesNoahNuSmallerThanGamma}
\end{equation}

In this case, we associate $W(z_\beta)$, Eq. \eqref{ICD}, now with the infinite-invariant density \eq{\mathcal{I}(v)}, Eq. \eqref{VelocityID}, and $\alpha=2L-1, \beta=0$, so $z_\beta=v$. The Noah effect tells us that the asymptotic shape of the increment PDF is given by a non-normalizable function, and the relation between $M$ and $L$ here also agrees with Eq. \eqref{LAndMID}, as it should.   
Fig. \ref{FigICD}c shows how simulation results of $t^{\alpha}P_t(v)$ at converge increasing times to $\mathcal{I}(v)$, the infinite invariant density. 

{As in the other regimes, here the mean duration of the L\'evy walk steps in \eq{g(\tau)}, Eq. \eqref{MLDistributionOfTaus} is divergent, however since $\nu$ is small, the step velocity decays very quickly with the duration. Therefore the step displacement $\chi\sim\tau^\nu$, is almost decoupled from $\tau$.
This implies too things: First, the MSD of the process now mostly depends on how many steps the walker can have between $t'=0$ and $t$, and that is determined only by the value of $\gamma$. So the Hurst exponent in this regime depends only on $\gamma$.  Second, by a hand-waving argument we can see why $M$ and $L$ depend only $\gamma$; because if the step displacement depends only on this parameter, the average velocity $\MeanV$ in all the time-series increments withing those steps will depend only on this parameter too.
\begin{figure*}
\centering
\includegraphics[width=1\textwidth]{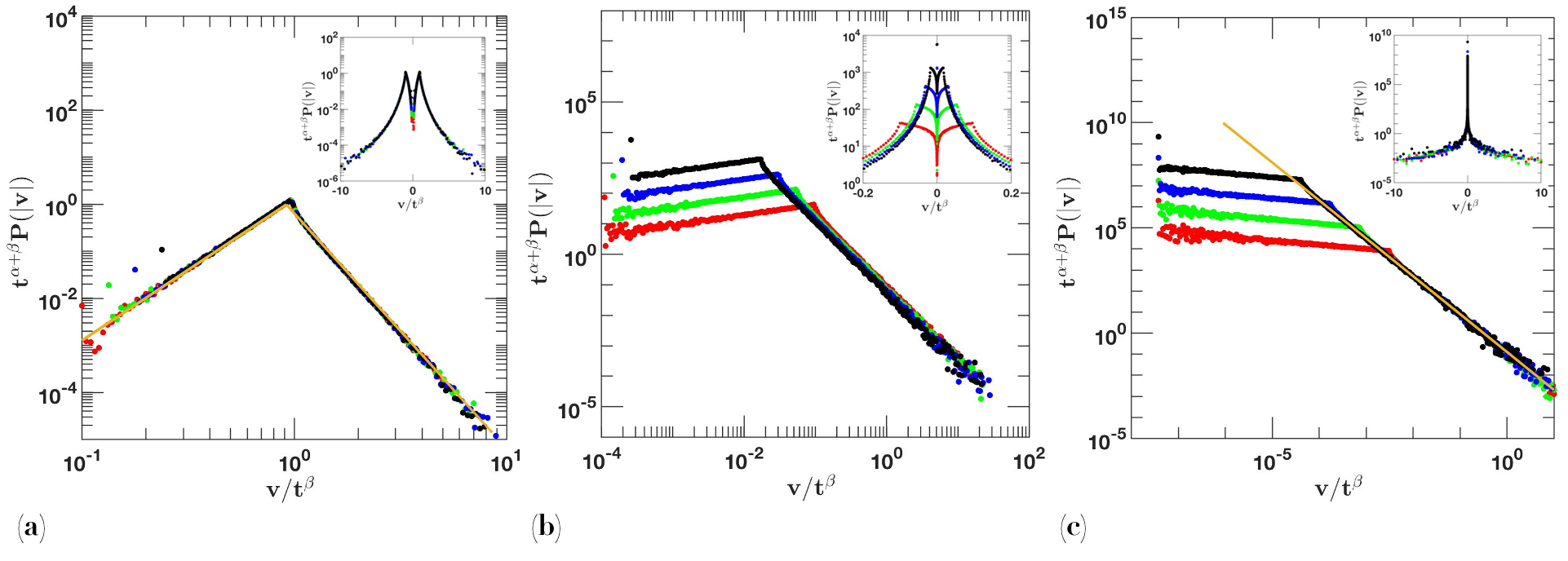}
\caption {\footnotesize{Numerical examination of the convergence of the increment PDF $P_t(\MeanV)$ to a time-invariant shape, based on Eqs. (\ref{ICD},\ref{ICD1}) and the quantification of the Moses and Noah effects. From Eq. \eqref{ICD1}: $\alpha = 2L - 1$ and $\beta = M + \alpha - 0.5$. In the Log-log plots (a),(b), and (c), in symbols, we see the rescaled PDF $t^{\alpha+\beta}P_t(\MeanV)$, obtained from simulation results of $3*10^8$ paths, in regimes A,B and D respectively (in regime C the shape of the PDF is behave similar to the last, at increasing times). The measurements were performed at times $t = 10^4$ (red dots), $10^5$ (green dots), $10^6$ (blue dots) and $10^7$ (black dots). (a) Here we used $\gamma = 0.5$, $\nu = 0.875$, leading to $L = 0.5$, and $M = 0.375$. The figure shows that the simulation results converge at increasing times to the normalized scaling shape given in Eq. \eqref{WzBetaSmallV} (solid mustard line). (b) Here,  $\gamma = 0.5$, $\nu = 0.625$, and $L = 0.625$, $M = 0.125$. Attempting to find an asymptotic scaling shape in this regime, which corresponds to Eq. \eqref{ICD}, does not work, since $\langle|\MeanV|\rangle$ and $\langle\MeanV^2\rangle$ do not correspond to a single scaling regime of $P_t(\MeanV)$. (c) Here $\gamma = 0.5$, $\nu = 0.375$, and $L = 0.75$, $M = 0$. $M$ and $L$ here obey the scaling relation in Eq. \eqref{LAndMID}, hence we expect to find that the increment PDF approaches the shape of a non-normalizable infinite-invariant density. This is confirmed by the solid mustard line, that represents  Eq. \eqref{VelocityID}. The insets show the same results, but in semi-log plots.}}
 \label{FigICD}
\end{figure*}

\section{\eq{J}, and how we obtained it}
\label{JosephExponent} 

The Joseph exponent depends on the shape of the auto-correlation function. However, this quantity is difficult to obtain for many systems, analytically and numerically.
In practice, the Joseph effect is often quantified  by designated methods, such as the so-called rescaled range statistic (R/S) \cite{Hurst}, wavelet decomposition \cite{WAVELET} or detrended fluctuations analysis \cite{Pen94}. Additional information on the correspondence between our definition of $J$ and the latter method is given in appendix \ref{DFAAndEATMSD}.

Here, for the L\'evy process, we use a measure which is easier to handle analytically; the ensemble averaged time-averaged MSD  $\left\langle\overline{\delta^2}\right\rangle$, defined as    
\begin{equation}
\langle \overline{\delta^2} \rangle \equiv{} \left\langle{} \frac{1}{t-\Delta} \int_0 ^{t-\Delta} 
 \left[ x(t_0+\Delta) - x(t_0)\right]^2  \mbox{ d} t_0\right\rangle{}. 
 \label{EATAMainText}
\end{equation} 
Note that Eq. \eqref{EATAMainText} should not be confused with \eq{\langle\overline{\MeanV^2}\rangle} in Eq. \eqref{NoahDefinitionContinousTime}, since in the latter the increments are strictly non-overlapping, whereas in \eq{\langle\overline{\delta^2}\rangle} they are.  
This quantity is related to the auto-correlation function, via  \cite{meyer2017greenkubo}
\begin{equation}
    \langle \overline{\delta^2} \rangle \approx \frac{2}{t}\int_0 ^{t}\!\!\mathrm{d}t_0\int_0^{\Delta}\!\! \mathrm{d}t_2\int_0^{t_2}\!\! \mathrm{d}t_1 \langle v(t_1+t_0)v(t_2+t_0) \rangle,
    \label{xmxofC}
\end{equation}
when $t\gg \Delta$. The scaling of this function for different types of auto-correlations is discussed in Appen. \ref{Appenctamsd}, where we also show the correspondence between $\langle\overline{\delta^2}\rangle$, the autocorrelation function, and our definition in Eq.  \eqref{JosephDefinitionContinousTime}. In all the cases considered in the appendix (even for $J\leq 1/2$), the asymptotic scaling is
\begin{equation}
    \langle \overline{\delta^2} \rangle \sim t^{2L+2M-2} \Delta^{2J}. 
    \label{EATAMSDScaling}
\end{equation} 
Our model is described by type (II) in the appendix.
This means that the Joseph exponent is given by the scaling of $\langle \overline{\delta^2} \rangle$ with the lag time $\Delta$ (note, that this observation was already made in \cite{meyer2018anomalous}, however, due to a typo it was read '$t$' instead of '$\Delta$').

To obtain the Joseph exponent in various regimes, we use the results of the calculation of the ensemble-time averaged MSD, obtained for this model in Ref. \cite{albers2018exact}. 
Particularly, in that Ref., the scaling of $\langle \overline{\delta^2} \rangle$ with respect to time and \eq{\Delta} was calculated for a general shape of \eq{g(\tau)}, with an asymptotic fall-off as in Eq. \eqref{MLDistributionOfTaus}, at large \eq{\tau}, and it was shown to not depend on the exact behavior at small \eq{\tau}s. 
Given this knowledge, the time-averaged MSD (for \eq{0<\left\{\gamma,\nu\right\}<1)} for the L\'evy walk model we study here, has the following scaling \cite{albers2018exact}
\begin{align}
    \langle \overline{\delta^2}\rangle\propto\begin{cases} 
    t^{2\nu-2}\Delta^2,\qquad\qquad \HHquad \gamma/2+1/2<\nu \\ 
    t^{\gamma-1}\Delta^{1+2\nu-\gamma},\qquad \gamma/2<\nu<\gamma/2+1/2\\
     t^{\gamma-1}\Delta,\qquad\qquad\quad \nu<\gamma/2
    \end{cases}.
    \label{TAMSDRadons}
\end{align}
Using Eq. \eqref{TAMSDRadons} and Eq. \eqref{EATAMSDScaling}, we find that \begin{align}
    J=\begin{cases} 
    1,\qquad\qquad\qquad\qquad\quad\quad \gamma/2+1/2<\nu\\ 
    (1+2\nu-\gamma)/2,\qquad\quad\quad\gamma/2<\nu<\gamma/2+1/2\\
    1/2,\qquad\qquad\qquad\qquad\quad \nu<\gamma/2
    \end{cases}.
    \label{jRadons}
\end{align}

Note that since $\gamma<1$, the mean step duration in all these regimes diverges. This implies that the random walker essentially walks in the same direction for almost all of the time $t$, regardless how long it is. In turn, this means the process is correlated in the whole parameter regime that we study. But when $
\nu<\gamma/2$, the average correlations decay rather quickly with $\Delta$ because the step velocity changes only very little with the step duration, hence in this regime we do not see a Joseph effect (the difference between the mean velocity at increments belonging to the same steps of the L\'evy walk, versus increments of other steps, is small). The onset of the effect is above the line $\nu=\gamma/2$. It is maximal when $J=1$, at $\gamma+1<2\nu$.  

Fig. \ref{FigOfFourPhaseDiagrams}c shows a phase diagram summarizing the different regimes of the Joseph effect, shown in Eq. \eqref{jRadons}. Fig. \ref{FigOfFourPhaseDiagrams}d shows the different regimes of the Hurst exponents which results from the combined affect of the various effects leading to the anomalous diffusion, and calculated using Eq. \eqref{hjlm}. Our simulation results for several arbitrary samples of values of $\nu$ and $\gamma$ in these regimes agree with the analytic expectation. 
\section{A Generalized model} 
 \label{Section5} 
In this section, following \cite{albers2018exact,bothe2019mean} we extend the model displayed above by introducing a new parameter $\eta$. This parameter generalize Eq. \eqref{NonlinearDurationVelocityCoupling} by modifying the relation between the $i$th step velocity $V_i$, its duration $\tau_i$ and the actual time in motion $t'$, as follows 
\begin{equation}
    V_{\nu,\eta} = \pm\tilde{c}_1 \tau^{\nu-\eta} t'^{\eta-1}. 
\end{equation}
Some values of $\eta$ correspond to special cases: The L\'evy walk we studied above corresponds to $\eta=1$, when  $\eta=\nu$ we get a Drude-like model \cite{schulz1997anomalous,benkadda1998chaos}, and when  $\eta \rightarrow 0$ or $\eta\rightarrow\infty$ we approach either a jump-then-wait type of coupled continuous-time random walk, or a wait-then-jump model, respectively  \cite{bothe2019mean}. As we will now show, modifying this parameter changes the onset of the ``$\infty$" regime. Our simulation results suggest that when $\eta$ is within the open range $(0,\infty)$, the behavior of all the effects in regimes A,B,C,D in Fig. \ref{fig2:QgphaseAllTogether} does not change, however the regimes themselves may  expand or shrink and disappear. 

Let's look again at the PDF {$P_t(x)$}, of the particles' displacement $x$ at time $t$. Here,  
\begin{align}
    P_t(x) = \int_{-\infty}^{\infty} \Intd x' \int_{0}^{t} \Intd t' A(x',t') r(x-x'|t-t'). 
    \label{maineqP}
\end{align}
 where $A(x',t')$ is the joint probability density to land on $x'$ between $x$ and $x+dx$ in a complete step ending at $t'<t$, and $r(x-x'|t-t')$ is the conditional probability density of the displacement in the last, incomplete step given the duration of the walk is $t$. 
 The following calculation of the MSD for this model is adapted from Ref. \cite{bothe2019mean}. 
 Let  $\hat{f}(k,t)=\int_{-\infty}^\infty f(x,t)\exp(-i kx)\Intd x$, be the Fourier transform of some function $f(x,t)$, from $x\rightarrow k$. Eqs. (\ref{FourierOfP},\ref{FourierOfr}) and Eq. (\ref{FourierOfA}) below, represent the characteristic functions of the probability densities $P,r$ and $A$, respectively. All the functions except for $P$ lack normalization on unity; the zero terms of the expansions are denoted as $r_0(t) = \int r(x|t) dx \neq 1$ and $A_0(t) = \int A(x,t) dx \neq 1$. Let $x_2(t) \equiv \langle x^2(t) \rangle$ be MSD at time $t$, $A_2(\tau) = \int \chi^2A(\chi,\tau) \Intd \chi$ is the marginal second moment of displacement $\chi$ in a single complete step of duration $\tau$ and $r_2(\tau^*) \equiv \int {\chi^*}^2r(\chi^*|\tau^*) \Intd \chi^*$ is the MSD of the displacement $\chi^*$ in the last, incomplete step. The duration $\tau^*$ of the latter is defined in Eq. \eqref{TotalTime}. After Fourier transform, we get
 \begin{equation}
       \hat{P}(k|t) = 1 - \frac{1}{2} k^2x_2(t) + o(k^2) 
       \label{FourierOfP}
 \end{equation}
\begin{equation}
\hat{r}(k|t) = r_0(t) -\frac{1}{2}k^2r_2(t) + o(k^2)
\label{FourierOfr}
\end{equation}
\begin{equation}
    \hat{A}(k|t) = A_0(t) - \frac{1}{2}k^2A_2(t) + o(k^2)
    \label{FourierOfA}
\end{equation} 
 Let $\hat{f}(k,s)=\int_{0}^\infty \hat{f}(k,t)\exp(-st)\Intd t$ be the Laplace transform of $\hat{f}(k,t)$. In Fourier and Laplace space, from Eq. \eqref{maineqP} we obtain 
\begin{equation}
\small
    \hat{P}(k,s) = A_0(s)r
    _0(s) - \frac{k^2}{2}\left[A_0(s)r_2(s) + A_2(s)r_0(s)\right] + o(k^2)
    \normalsize
    \label{FourierLaplaceOfP}
\end{equation}

 On comparing Eq. \eqref{FourierOfP} and Eq. \eqref{FourierLaplaceOfP}, we can now obtain the MSD using 
 \begin{equation}
     \langle x^2(s) \rangle = A_0(s)r_2(s) + A_2(s)r_0(s)
     \label{LaplaceOfx}
 \end{equation}
 Calculating the values on the right-hand side of Eq. \eqref{LaplaceOfx}, and taking  the inverse Laplace transform, one can now derive the MSD. Part of this calculation, performed in \cite{bothe2019mean}, was to obtain the second marginal moment of the function $r(x|t)$:
 \begin{align}
     r_2(t) \simeq \gamma\tilde{c}_1^{2}\tau_0^{\gamma}t^{2\eta} \int_{t}^{\infty} t'^{2(\nu-\eta)-1-\gamma}{d}t'.
 \end{align}
 From here, we can see that $r_2$, and therefore also $\langle x^2\rangle$, can only obtain a finite value when $\gamma > 2(\nu-\eta)$. This explains the crossover to the ``$\infty$" regime, which occurs when $\nu>\gamma/2+\eta$. 
 
 When $\nu<\gamma/2+\eta$ and $\gamma < 1, 2\nu < \gamma$, the MSD is \cite{bothe2019mean}  
\begin{align} 
\small 
    \langle x^2(t)\rangle \approx \gamma &\left[ \frac{\Gamma(2\nu+1-\gamma)}{\Gamma(1-\gamma)(2(\nu-\eta)-\gamma)\Gamma(2\nu+1)} t^{2\nu}\right.\nonumber\\ 
    &\left.+ \frac{B(2\nu+1,\gamma-2\nu)}{\Gamma(1-\gamma)\Gamma(1+\gamma)}\tilde{c}_1^{2} t^{\gamma} \right], 
\normalsize 
\end{align}
where $B(a,b)$ is the Beta-function. This is dominated by the second term, since $2\nu < \gamma$, and therefore $\langle x^2(t) \rangle \propto t^{\gamma}$, which gives the value of the Hurst Exponent as $H=\gamma/2$, similar to what is seen seen in Fig. \ref{FigOfFourPhaseDiagrams}d. When $\nu<\gamma/2+\eta$, but  $\gamma <1, 2\nu > \gamma$, the MSD reads \cite{bothe2019mean} 
\begin{align}
\small
    \langle x^2(t)\rangle \simeq \gamma\frac{\Gamma(2\nu-\gamma)}{\Gamma(2\nu+1)\Gamma(1-\gamma)}\frac{4\nu-2\eta-2\gamma}{2(\nu-\eta)-\gamma} \tilde{c}_1^{2}t^{2\nu}, 
\normalsize
\end{align}
which gives the value of the Hurst Exponent as $H=\nu$ also similar to Fig. \ref{FigOfFourPhaseDiagrams}(d). The results shown in Fig. \ref{FigOfFourPhaseDiagrams} are for $\eta = 1$, whereas Eqs. (\ref{FourierLaplaceOfP},\ref{LaplaceOfx}) are calculated for any value of $\eta$. This shows that the power-law dependence of the mean squared displacement is independent of the exponent $\eta$ and the particular value of $\eta$ only enters in the prefactors, when $\langle x^2\rangle$ is finite. When $\eta$ is very small, only regime D in Fig. \ref{fig2:QgphaseAllTogether} survives, and beyond it we have the non-scaling ``$\infty$" regime. When $\eta$ is very large, regime A expends higher into the realm of $\nu>1$. 

\section{Discussion} 
\label{Discussion} 

Imagine that you get hold of a ``blind" set of data series, containing the positions of an ensemble of random walkers at various times in the interval \eq{[0,t]}. This data was generated by a L\'evy walk model, but you do not have this prior knowledge. Our analysis allows us to uncover the main features of the hidden process that cause its behavior to scale anomalously with time, despite the fact that we do not know what process generated the data. Elucidating the origins of anomalous diffusion observed in experimental  data is crucial in order to understand the underlying functioning of the system, and it is studied therefore these days , e.g., using new advanced methods for single-particle tracing  \cite{kepten2011ergodicity,weigel2011ergodic,wang2017three,sabri2020elucidating}. We encourage the verification of our results for example in (but not limited to) future such experiments, in particular e.g. the scaling relation in Eq. \eqref{hjlm} and Eq. \eqref{ICD}, and consequently its application. 

In addition to learning about the origins of the anomalous diffusion, one may use the knowledge about the Moses, Noah and Joseph effects in order to try to extrapolate which processes can and cannot be at least good candidates to represent the underlying dynamics. These days, there are many studies which use techniques such as machine learning  \cite{munoz2019machine,bo2019measurement,janczura2020machine}, Bayesian statistics \cite{thapa2018bayesian} and more, e.g.  \cite{kepten2013improved,thapa2020leveraging}, to try to infer the Hurst exponent or distinguish between various known models such as continuous-time random walk, fractional Brownian motion and others, which lead to anomalous scaling of the MSD, based only on analysis of data obtained from single trajectories. 
This issue is even being studied today as part of a multi-group competition to characterize the properties of anomalous diffusion in data, called the ANDI challange \cite{munoz2020andi}. 
Though we cannot fully and uniquely restore the underlying dynamics just by discerning the scaling properties of the process from the data, the characterization of anomalous diffusion using three additional exponents $M,L$ and $J$, in addition to the Hurst, does bring additional tools which can be helpful for modeling it. In this sense, this decomposition should also be useful for example for the modeling of diffusion in the membranes of living cells done in \cite{weigel2011ergodic}, where a Moses and a Joseph effect seem to have been observed. Another interesting example is found in  \cite{tabei2013intracellular}, where the authors observed  intercellular transport of insulin granules in eukrayotic cells, and then used information from the time-averaged MSD (Joseph), and the evolution of the absolute-mean of the increments (Moses) to model it. The authors compared two candidate models to describe their dynamics: fractional Brownian motion and continuous-time random walk, and concluded that non is sufficiently good for a full description of the system. They therefore continued by proposing a different, `hybrid' model based on the previous two \cite{tabei2013intracellular}. Since the first model leads only a Joseph effect, but the second leads to both Moses and Noah, a full three-effect decomposition here, which takes into account also the inherent relation between them, Eq. \eqref{hjlm}, might shed more light on the unified model. Of-course, in any  case if one seeks to fully reconstruct the underlying process from the data, a complete knowledge of the entire correlation structure would be required, including two-point, and all the higher order correlations. 

The L\'evy walk model that we studied in this paper, is a prototypical example which shows how the three effects analysis can be used for many other processes as well. The results in Fig. \ref{fig2:QgphaseAllTogether} and Fig. \ref{FigOfFourPhaseDiagrams} eventually only depend on two inputs: the shape of the step durations PDF at large $\tau$s, and the coupling between the step durations and the velocity, which can also be translated to the coupling between the step  duration and displacement, since the step-displacement $\chi=V\tau$. Therefore, a class of process which can be mapped into a coupled step-duration and step-displacement process, which also includes other processes such as the Pommeau-Manneville map \cite{meyer2018anomalous} and ATTM \cite{massignan2014nonergodic}, will display similar properties as in the various regimes in Fig. \ref{fig2:QgphaseAllTogether} and Fig. \ref{FigOfFourPhaseDiagrams}. These phase diagrams describe their dynamics as well, after change of variables (see e.g., \cite{meyer2018anomalous}). 
\\

\textbf{Acknowledgement}
\\
This research is supported by the Max-Planck society (EA, PGM and HK), and in part by the US National Science Foundation, through grant DMR-1507371 (VA and KEB).  KEB also thanks the Max-Planck-Institut für Physik komplexer Systeme for its support and hospitality during his visit when this work was initiated.

\appendix{


\section{Generality of $H=L+J+M-1$}
\label{AppenHVsMLJ}

As mentioned in the main text, the summation relation, Eq. \eqref{hjlm}, between $M,L,J$ and $H$ was previously presented for several examples of processes, in \cite{chen2017anomalous,meyer2018anomalous}. These studies suggest that this relation is also valid for a much larger range of systems, even though a unified proof is still required in a future work. The following example, shows that we can prove this relation analytically also for a widely useful system where $J\leq1/2$ (in the derivation  in Eq. \eqref{ConnectionBetweenExponents1} we assumed that $J>1/2$), and the correlation function is negative. 
Consider the ARFIMA$(0,d,0)$ process \cite{Beran,Wat17} $x(t)=\sum_{i=0}^t X_i$, in discrete time, whose increments are defined via the  transformation 
$(1-\hat{B})^d X_i=\sigma^2\eta(i)$, where $\eta_i$ is Gaussian white noise with zero mean and $\langle\eta_i\eta_j\rangle=\delta_{ij}$. Here,  $\hat{B}^nX_i=X_{i-n}$, $d<0$ and $(1-B)^d=\sum_{k=0}^\infty\frac{\Pi_{a=0}^{k-1}(d-a)(-B)^k}{k!}=1-dB+\frac{d(d-1)}{2!}B^2+...$. When $-1/2\leq d\leq 0$, this process is long-ranged anti-correlated, and the autocorrelation function of the increments of this process is~\cite{Beran}: 
\begin{equation} 
\small
c(\tilde{\Delta})= \langle X_{i+\tilde{\Delta}}X_i\rangle=\prod_{k=1}^{\tilde{\Delta}} \frac{k-1+d}{k-d}= \frac{\Gamma(\tilde{\Delta}+d)\Gamma(1-d)}{\Gamma(\tilde{\Delta}-d+1)\Gamma(d)}.  
\normalsize 
\label{ARFIMACorrelationFunction} 
\end{equation} 
For large $\tilde{\Delta}$; $c(\tilde{\Delta})\sim\frac{\Gamma(1-d)}{\Gamma(d)}\tilde{\Delta}^{2d-1}$, so according to the definition in Eq. \eqref{JosephDefinitionContinousTime}, the Joseph exponent is $J=d+1/2$.
The MSD is related to the correlation function via 
\begin{align}
    \langle x^2(t)\rangle &= \langle (\sum_{i=1}^t x_i)^2\rangle= \sum_{i,j=1}^t\langle x_i x_j\rangle\nonumber\\ 
    & = \sigma^2( t + 2\sum_{\tilde{\Delta}=1}^t (t-\tilde{\Delta}) c(\tilde{\Delta})).
    \label{ARFIMAMSD}
\end{align}
Plugging Eq. \eqref{ARFIMACorrelationFunction} into Eq. \eqref{ARFIMAMSD}, we find that at long $t$ 
\EQ{\langle x^2(t)\rangle\sim \frac{t^{2 d+1} \left| \Gamma (-d)\right| }{(2 d+1) \Gamma (d)}+\frac{d}{2 d+1}+\mathcal{O}\left(t^{2 d-1}\right),}{MSDARFIMALargeT}
and from the leading-order term, using  $J=d-1/2$, we find $\langle x^2(t)\rangle\propto t^{2J}$, namely $H=J$. This is a well known result, and note that in the standard ARFIMA process the increment distribution is stationary and thin-tailed, hence $M=L=1/2$ and the summation relation in the section title, and Eq. \eqref{hjlm} is fulfilled. Now, consider the related process: $\tilde{x}(t)=\sum_{i=0}^t\tilde{X}$, where $(1-\hat{B})^d X_i=\sigma^2t^{2L+2M-2}\eta(i)$. Here we introduced the time dependence of the variance of the increments in the same way is in Eq. \eqref{NoahDefinitionContinousTime}, which means that now the process can have both a Noah and a Moses effect, in addition to Joseph.  A calculation in this case, which follows exactly the same lines as the above, will now yield $\langle x^2(t)\rangle\sim t^{2L+2M+2J-2}$, which again leads to Eq. \eqref{hjlm}. 
Further generalizations of this summation relation are discussed below, in Appen. \ref{Appenctamsd}.


\section{The ensemble-time averaged MSD and the correlation function}
\label{Appenctamsd}

In this section we want to clarify the connection between the autocorrelation function and the Ensemble-Time (ensemble averaged - time averaged) Mean-Squared Displacement (EATA MSD) and thereby show the types of correlations that lead to specific values of the Joseph exponent $J$.
The time-averaged MSD defined in Eq.  (\ref{EATAMainText}) depends on two times,  $t$ and $\Delta$. Here $t$ is the measurement time and $\Delta$ is the lag time. However, for some systems, especially the ones of interest for this study, the time-averaged MSD does not converge to a single value. For an analytical approach, we therefore consider the ensemble average of the time-averaged MSD, namely the EATA MSD
\begin{equation}
\label{TAMSDdivide}
\langle \overline{\delta^2} \rangle \approx  \frac{1}{t} \int_0 ^{t} 
\langle \left[ x(t_0+\Delta) - x(t_0)\right]^2\rangle  { d} t_0.
\end{equation}
As discussed in \cite{meyer2017greenkubo}, in the limit $t\rightarrow \infty$
only the upper bound of the integral is important and the behavior around zero is negligible. 

Now for the expression below the integral we have to find the  MSD recorded between 
time $t_0$ and $t_0+\Delta$ under the condition that $t_0\gg\Delta$.
It is given in equation (\ref{ConnectionBetweenExponents1}).
It connects the integrand in Eq. (\ref{TAMSDdivide}) to the velocity correlation function. 
We denote
\begin{equation}
    \langle \left[ x(t_0 + \Delta) - x(t_0)\right]^2 \rangle = 2\int_0^{\Delta}\!\! dt_2\int_0^{t_2}\!\! dt_1 \langle v(t_1+t_0)v(t_2+t_0) \rangle.
    \label{xmxofC}
\end{equation}
Now we want to discuss four different cases, i.e. three different types of correlation functions.

(I) In the first step we consider correlation functions $\langle v(t_1+t_0)v(t_2+t_0) \rangle=C(t_2-t_1)$, that do not depend on the measurement time $t_0$. 
In this case the only way to violate the Gaussian CLT is with diverging correlation times, i.e. the correlation function asymptotically scales like a power law
\begin{equation}
    \langle v(t_1+t_0)v(t_2+t_0) \rangle \sim (t_2-t_1)^{2J-2}.
\end{equation}
The equivalence of $H$ and $J$ can also be found by plugging the correlation function into equation (\ref{ConnectionBetweenExponents1}).
Now using equation (\ref{xmxofC}), the exact same scaling, 
$\langle \left[ x(t_0 + \Delta) - x(t_0)\right]^2 \rangle \sim \Delta^{2H}$, 
is obtained. 
Since this still does not depend on $t$, the EATA MSD exhibits the same scaling
\begin{equation}
    \langle \overline{\delta^2} \rangle \sim \Delta^{2J}.
    \label{TAMSD1}
\end{equation}

(II) For the second scenario we consider a  correlation functions that, asymptotically, exhibits power-law scaling, both in $t$ and $\Delta$. 
Now, we can write the correlation function as
\begin{equation}
    \langle v(t) v(t+\Delta) \rangle \sim t^{2H-2} \Phi\left( {\Delta \over t} \right) ,
\end{equation}
where $\Phi(q)$ (and $q=\Delta/t$), is a positive valued function describing asymptotic scaling.
The scaling exponent $2H-2$ can again be obtained using Eq.  (\ref{ConnectionBetweenExponents1}). This was shown in \cite{Dechant14}. The total measurement time in this case has to be much larger than the lag time $\Delta \ll t$, then only the small-$q$ asymptotic behavior of $\phi(q)$ is relevant
\begin{equation}
    \phi ({q}) \sim {q}^{ 2J-2} \ \ \mbox{with} \ \  2-2H\le 2-2J <1 \ \ \ \  {q} \rightarrow 0.
\end{equation}
The conditions above are necessary in order to ensure that the correlation function decays with $\Delta$, and at the same time does not blow up with time $t$.
Since in the $q=0$ case, the correlation function is equal to the velocity displacement Eq. (\ref{NoahDefinitionContinousTime}), continuity demands
\begin{equation}
2-2J = 2L+2M - 2H \Leftrightarrow H=J+L+M-1 .
\end{equation}
Now, the correlation function for $q\rightarrow 0$ can be inserted into Eq.  (\ref{xmxofC})
\begin{align}
\label{eq17a}
&\langle \left[ x(t_0 + \Delta) - x(t_0)\right]^2 \rangle \sim \nonumber\\ & \ \ \ \ \ \ \ \ 
2
\int_0 ^{\Delta}\! { d} t_2 \int_0 ^{t_2}\! { d} t_1
\left( t_1 + t_0\right)^{2L+2M-2} 
\left( {{t_2 - t_1 }} \right)^{2J-2}.
\end{align}
Integration yields with $t_0\gg t_1, t_2$
\begin{align}
\label{eq17a1}
&\langle \left[ x(t_0 + \Delta) - x(t_0)\right]^2 \rangle \sim \nonumber\\ & 
\frac{2}{\left(2J-1 \right) \left(2J \right) }
(t_0)^{2L+2M-2} \Delta^{2J}. 
\end{align}
Inserting this result into Eq.  (\ref{TAMSDdivide}) only yields an additional pre-factor. Therefore, the EATA MSD scales like
\begin{equation}
    \langle \overline{\delta^2} \rangle \sim t^{2L+2M-2} \Delta^{2J}.
    \label{TAMSD2}
\end{equation}{}

(III) The third case occurs for processes with correlations that do scale with the measurement time $t$, but decay faster with the lag time $\Delta$. Here, we write the correlation function as
\begin{equation}
    \langle v(t) v(t+\Delta) \rangle \sim t^{2H-2} \Phi\left( {\Delta} \right) ,
    \label{CIII}
\end{equation}
with a positive function $\Phi(\Delta)$, which decays faster than $\Delta^{-1}$, i.e. the integral over the autocorrelation function with respect to $\Delta$ becomes finite for $\Delta\rightarrow\infty$. 
The dependence on $H$ can again be verified using Eq. (\ref{ConnectionBetweenExponents1}).
Here, we specify the shape to be $\Phi\left( {\Delta} \right)={(1+\Delta)}^{2\varphi -2}$ with $\varphi<1/2$. 
The calculation is as simple for the relevant cases of exponential decay or the velocity correlation function being (Dirac-) delta-distributed with $\delta(\Delta)$.
Using Eqs. (\ref{xmxofC}) and (\ref{CIII}) we find for $t_0\gg t_1,t_2$
\begin{align}
\label{src}
&\langle \left[ x(t_0 + \Delta) - x(t_0)\right]^2 \rangle\approx  \nonumber\\ & \ \ \ \ \ \ \ \ 
2 \int_0 ^{\Delta}\! { d} t_2 \int_0 ^{t_2}\! { d} t_1
\left( t_1 + t_0\right)^{2-2H} {(1+t_2 - t_1)}^{2\varphi -2}  \nonumber\\ & \ \ \ \ \ \ \ \ 
\approx \frac{2}{2\varphi -1}\int_0 ^{\Delta}\! { d} t_2 
\left( t_0\right)^{2-2H} \left({1}-{(1+t_2)}^{2\varphi -1}\right)  \nonumber\\ & \ \ \ \ \ \ \ 
\approx \frac{2}{2\varphi -1}(t_0)^{2-2H} \left(\Delta+\frac{1}{2\varphi}(\Delta^{2\varphi}-1) \right).
\end{align}
Since $\varphi<1/2$, for $\Delta\rightarrow\infty$, the linear term is dominant.
Accordingly, the scaling of the EATA MSD for short range correlated processes is
\begin{equation}
    \langle \overline{\delta^2} \rangle \sim t^{2-2H} \Delta.
    \label{TAMSD2}
\end{equation}{}
The Hurst exponent is independent of $\varphi$ and of the exact shape of the distribution as long as it decays sufficiently fast. The Joseph exponent is therefore always $J=1/2$.

(IV) In all cases discussed so far, the Joseph exponent is $J\geq 1/2$. This is also a necessary condition for the derivation in equation (\ref{ConnectionBetweenExponents1}). However, it is possible, to construct antipersistent processes with a autocorrelation functions, that lead to $J<1/2$. Looking at the calculation (\ref{src}) we see, that in order to obtain a scaling with $H<1/2$, the constant factor after the first integration, which leads to the linear scaling, has to be eliminated. This is possible if the correlation function is not strictly positive. The most prominent example of a continuous process, which can fulfill this condition is fractional Gaussian noise, which is the continuous-time version of the discrete ARFIMA$(0,d,0)$ studied in Appen. \ref{AppenHVsMLJ}. This process has a correlation function similar in shape as in case III above, but without the dependence on $t$, it reads
\begin{equation}
    \langle v(t)v(t+\Delta)\rangle=\frac{1}{2}\left(|\Delta+1|^{2J}-2|\Delta|^{2J}+|\Delta-1|^{2J}\right).
\end{equation}{}
Putting this into equation (\ref{src}) leads to $H=J<1/2$.

In conclusion we want to point out that the exponent $J$ is given by the scaling of the autocorrelation function with $\Delta^{2J-2}$ directly for $J>1/2$, and by the scaling of the integral over the autocorrelation function with $\Delta^{2J-1}$ for all cases.

\begin{figure*}
\centering
(a)\includegraphics[width=0.20\textwidth]{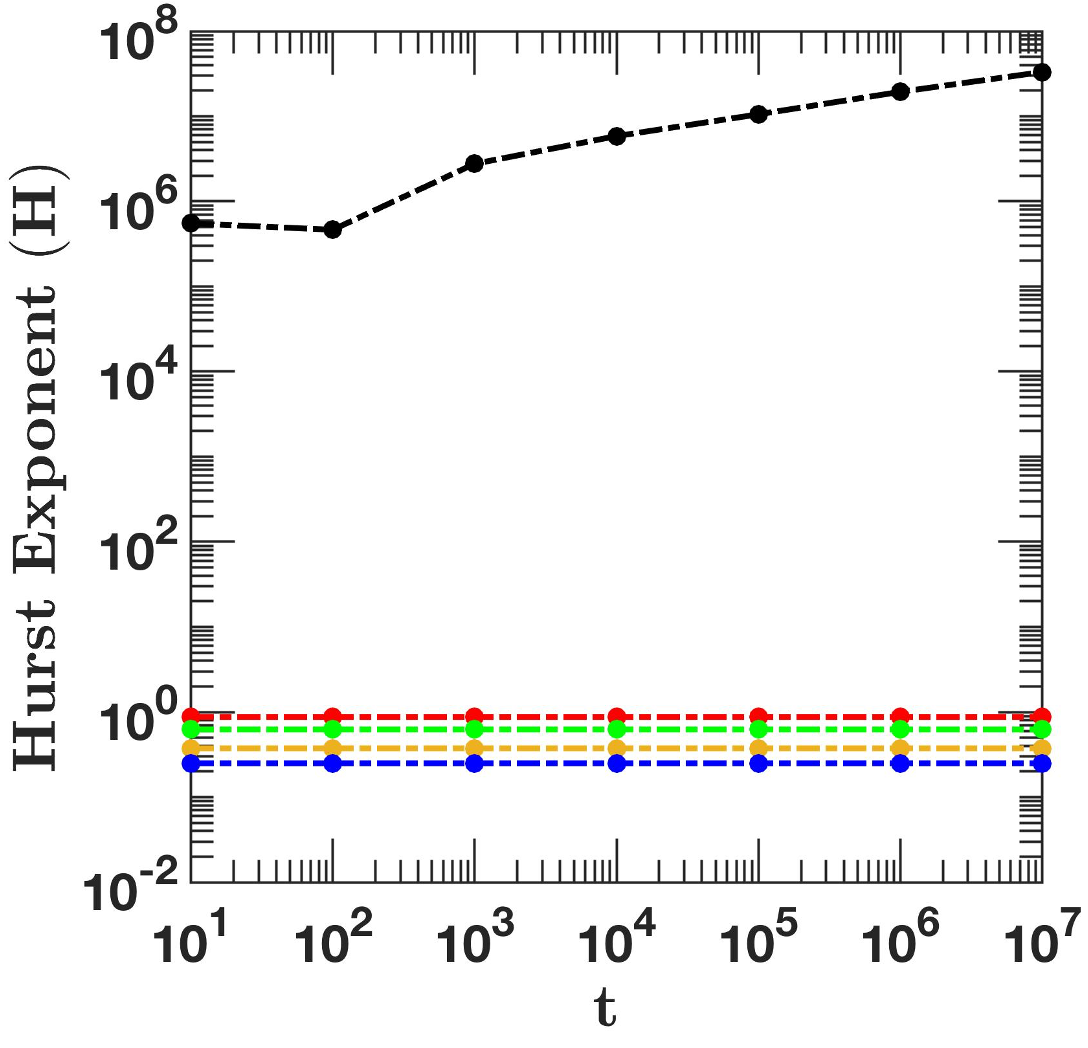}
(b)\includegraphics[width=0.20\textwidth]{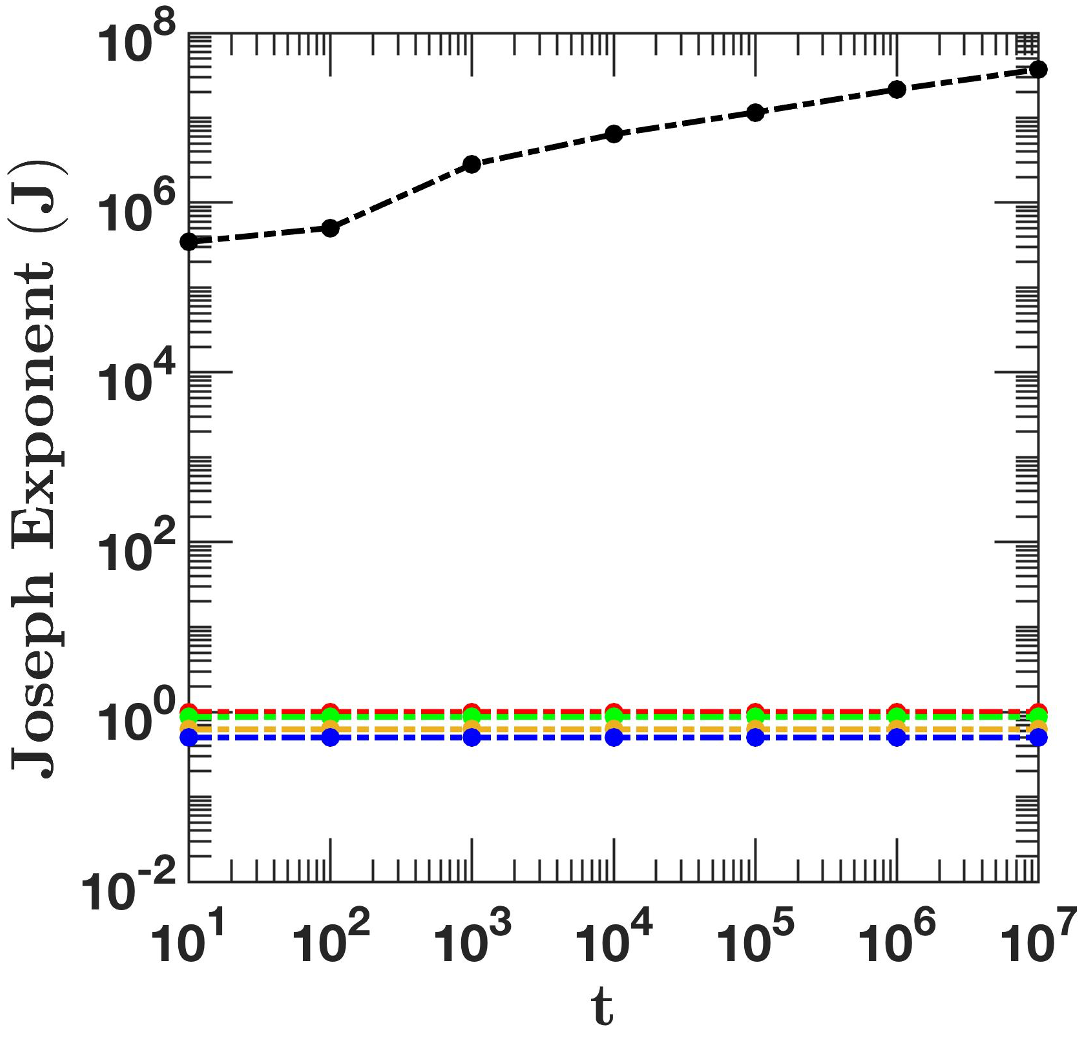}
(c)\includegraphics[width=0.20\textwidth]{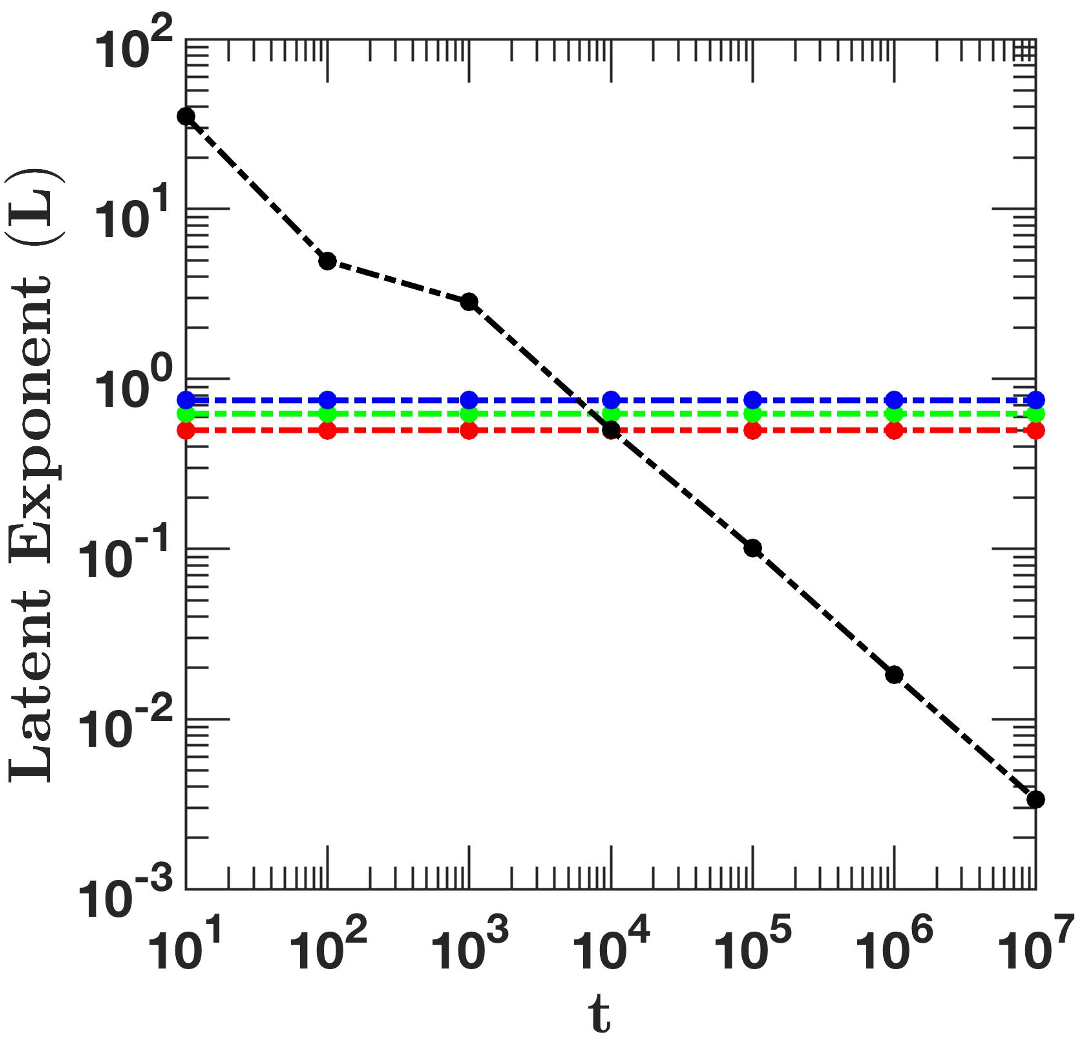}
(d)\includegraphics[width=0.20\textwidth]{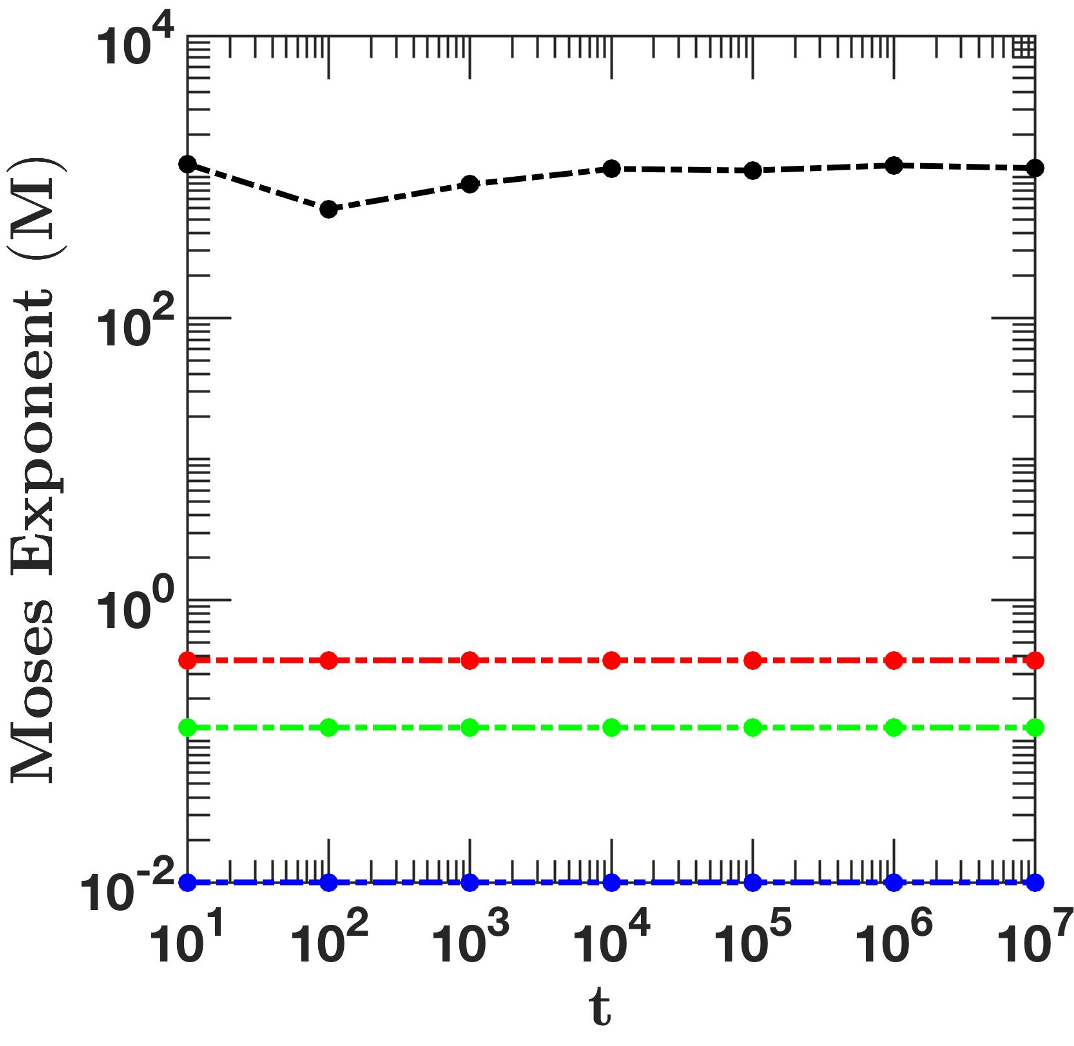}
 \caption {\footnotesize{ Log-log plots for the values of $\log(\langle{x^2}\rangle)/\log(t)$,~ $\log(\langle\overline{\delta^2}\rangle)/\log(\Delta)$,~$\left[\log\left(\frac{\sqrt{\langle\overline{v^2}\rangle}}{\langle\overline{|v|}\rangle}\right)/\log(t)+1/2\right]$ and $\left[\log(\langle\overline{|v|}\rangle)/\log(t)+1/2\right]$, which respectively represent the four exponents $H,J,L,M$ when they are constants, in different regimes. (a) The value of the Hurst exponent $H$, (b) Joseph exponent $J$, (c) Latent exponent $L$, and (d) Moses exponent $M$. The dashed red line represents the value of the exponents in region A for $\gamma = 0.5,\nu = 0.875$. The dashed green line represents the value of the exponents in region B for $\gamma = 0.5,\nu = 0.625$. The dashed mustard line represents the value of the exponents in region C for $\gamma = 0.5,\nu = 0.375$. The dashed blue line represents the value of the exponents in region D for $\gamma = 0.5,\nu = 0.2$. The dashed black line represents the value of the exponents in the $\infty$ region for $\gamma = 0.5$ and $\nu = 1.75$. In panels (a-c), the dashed black line does not converge to a constant, which means that in this region the Hurst, Joseph and Latent exponents are not well defined.}}
 \label{FigExopnentBeyondNu1}
\end{figure*}

\section{The Joseph effect via DFA}
\label{DFAAndEATMSD} 

There are several methods that are used in practice in order to quantify long range correlations in discrete measured time series. Examples are R/S statistics, detrended moving averages \cite{DMA}, scaling analysis based on the wavelet transform \cite{WAVELET} and detrended fluctuation analysis \cite{Pen94}. In \cite{Hoell19} it was shown that the latter three can be expressed in the same framework. In this section we want to discuss whether or not the definition of long range correlations in DFA is different from our definition, Eq.  (\ref{JosephDefinitionContinousTime}) for $J$.

The squared fluctuation function of DFA is defined as
\begin{equation}
    F_q^2(s)=\frac{1}{K}\sum_{k=1}^{K}\left( \frac{1}{s}\sum_{t=1+(k-1)s}^{k s} (x_t-p_{t,q})^2\right).
    \label{eq:DFAdef}
\end{equation}
Here $p_{t,q}$ is a polynomial that is fitted to the series $x_t$ for segments of lengths $s$. The squared error $(x_t-p_{t,q})^2$ of these fits is then averaged over all the non-overlapping segments of equal length $s$. The index $q$ is the order of the polynomial. DFA is not sensitive to trends with polynomial shape of order $q-1$, i.e. the slow dynamics is filtered by the method. For the definitions used above, data with trends does not yield a meaningful exponent. So we want to concentrate on stationary data.
If the detrending order is zero, in each window just the mean value is subtracted and Eq. (\ref{eq:DFAdef}) simplifies to a discrete version of the time-averaged MSD,  with non-overlapping windows.

So what is difference if detrending is performed with $q>0$? Here, for stationary systems, a relation between the fluctuation function and the autocorrelation function was derived in \cite{Hoe15}
\begin{equation}
    \langle F_q^2(s)\rangle = \langle v^2\rangle \left(L_q(0,s)+2\sum_{t=1}^{s-1}\frac{\langle v(t+\Delta)v(t)\rangle}{\langle v^2\rangle} L_q(t,s)\right).
\label{F_c}
\end{equation}
$L_q(0,s)$ is some sophisticated kernel. Its leading order is linear. So the fluctuation function is a measure of the integral (here discrete) over the autocorrelation function. Thus it measures the Joseph effect for $0<J<1$. 

If a Noah effect is present, i.e. the variance is infinite in theory, the formula can still be used, since a measured time series always has a finite variance \cite{meyerDFA}.
The Moses effect is more complicated. Even though a scaling of the increment distribution as in scaled Brownian motion (or in the parameter range A in the L\'evy walk) is not visible in DFA due to the averaging over the segments, the scaling exponent DFA still might differ from $J$. This is true if $v$ is diffusive as in fractional Brownian motion (DFA results shown in \cite{Hen00}). Here the DFA exponent is $>1$ in contrast to $J$. In fact it is equal to $H$.
So DFA is usually, but not always, a measure of the Joseph effect.
\\


\section{The ``$\infty$" regime, $\nu>\gamma/2+1$}
\label{AppendixInfinityRegime} 
As explained in Sec. \ref{Section5}, the analytic results show that when $\eta=1$ and $\nu>\gamma/2+1$, the Hurst exponent diverges. We demonstrate this in Fig. \ref{FigExopnentBeyondNu1}a, by showing simulation results for the time derivative of the ratio $\log(\langle x^2\rangle)/\log(t)$ for five pairs of $\nu,\gamma$ in the different regions shown in Fig. \ref{FigOfFourPhaseDiagrams}d. These simulation were performed with an ensemble size of $10^7$ trajectories. Clearly, in all but the ``$\infty$" regime, this ratio is constant. We expect to find a similar behaviour if we fix time and increase the ensemble size. 
In Fig. \ref{FigExopnentBeyondNu1}b-d, we checked the time evolution of $\log(\langle\overline{\delta^2}\rangle)/\log(\Delta)$, $\left[\log\left(\frac{\sqrt{\langle\overline{v^2}\rangle}}{\langle\overline{|v|}\rangle}\right)/\log(t)+1/2\right]$ and $\left[\log(\langle\overline{|v|}\rangle)/\log(t)+1/2\right]$ respectively, at increasing times, which yield the Joseph, Moses and Latent exponents when these ratios are constant (again, as seen in Fig. \ref{FigOfFourPhaseDiagrams}). We observe that in panels (b) and (c) of Fig. \ref{FigExopnentBeyondNu1} the above ratios are constant in all the regions except for the ``$\infty$'' regime. In the latter, the Joseph and the Latent exponents are clearly not well defined, which renders also the definition of $M$ at least irrelevant.  A similar behavior will appear also when $\eta\neq1$.

}

\bibliographystyle{aipnum4-1}
\bibliography{./bibliography2}

\end{document}